\documentclass[12pt]{article}
\pdfoutput=1

\usepackage{jheppub} 
\usepackage{graphicx}  
\usepackage{dcolumn}   
\usepackage{bm}        
\usepackage{amssymb}   
\usepackage{subcaption}
\usepackage{epstopdf}
\usepackage[dvipsnames]{xcolor}
\usepackage{slashed}
\usepackage[utf8]{inputenc}
\usepackage{multirow}
\usepackage{footnote}
\usepackage{amsmath}
\allowdisplaybreaks[4]
\usepackage{accents}
\newlength{\dhatheight}

\usepackage{hyperref}
\def\figureautorefname~#1\null{Fig.\,#1\null}
\def\tableautorefname~#1\null{Tab.\,#1\null}

\def\equationautorefname~#1\null{Eq.\,(#1)\null}

\definecolor{orange}{rgb}{1,0.5,0}
\definecolor{blue}{rgb}{0,0,1}

\title{Measuring Higgs Boson Self-couplings with $2\rightarrow 3$ VBS Processes}
\author[a,1]{Junmou Chen,}
\author[b,1]{Chih-Ting Lu,}
\author[c,1]{Yongcheng Wu}
\affiliation[a]{Department of Physics, Jinan University, Guangzhou, Guangdong Province, 510632, China}
\affiliation[b]{School of Physics, Korean Institute for Advanced Study, Seoul, 02455, Korea}
\affiliation[c]{Department of Physics, Oklahoma State University, Stillwater, OK, 74078, USA}

\note{First author.}

\emailAdd{chenjm@jnu.edu.cn}
\emailAdd{timluyu@kias.re.kr}
\emailAdd{ywu@okstate.edu}

\abstract{
     We study the measurement of Higgs boson self-couplings through $2\rightarrow 3$ vector boson scattering (VBS) processes in the framework of Standard Model effective field theory (SMEFT) at both proton and lepton colliders.
    The SMEFT contribution to the amplitude of the $2\to3$ VBS processes, taking $W_L W_L\rightarrow W_L W_L h$ and $W_L W_L\rightarrow h h h$ as examples, exhibits enhancement with the energy $\frac{\mathcal{A}^{\text{BSM}}}{\mathcal{A}^{\text{SM}}} \sim \frac{E^2}{\Lambda^2}$, which indicates the sensitivity of these processes to the related dimension-six operators in SMEFT.
    Simulation of the full processes at both hadron and lepton colliders with a variety of collision energies are performed to estimate the allowed region on $c_6$ and $c_{\Phi_1}$. Especially we find that, with the help of exclusively choosing longitudinal polarizations in the final states and suitable $p_T$ cuts,    $WWh$ process is as important as the more widely studied triple Higgs production ($hhh$) in the measurement of  Higgs self-couplings. Our analysis indicates that these processes can  play important roles in the measurement of Higgs self-couplings at future 100 TeV pp colliders and muon colliders. However, their cross sections are generally tiny at low energy machines, which makes them much more challenging to explore.}

\begin{document}

\maketitle
\newpage
\flushbottom

\section{Introduction}
\label{sec:intro}

The discovery of Higgs boson at LHC~\cite{Aad:2012tfa,Chatrchyan:2012ufa} marked a new era for particle physics. Although all experimental results so far agree with the standard model (SM),  the origin of electroweak symmetry breaking (EWSB) still remains a mystery and deserves more detailed studies~\cite{Higgs:1964pj,Englert:1964et,Guralnik:1964eu,Agrawal:2019bpm,DiMicco:2019ngk}.
Meanwhile the absence of BSM signal also makes precise measurements of Higgs properties more important than ever.
Especially, in order to determine the shape of the Higgs potential, the measurement of Higgs self-couplings becomes critical.

The most straightforward approach to measure Higgs couplings is through direct productions of Higgs boson(s). As the most notable example, the main channel to measure trilinear Higgs self-couplings at the LHC is the di-Higgs production through gluon-gluon fusion, see e.g. \cite{Sirunyan:2020xok,ATLAS:2021jki,Lu:2015jza,Adhikary:2017jtu,Goncalves:2018qas,Chang:2018uwu,Cepeda:2019klc,Cheung:2020xij} and references therein. However, processes related to longitudinal vector bosons can also be used for the measurement of Higgs couplings ~\cite{Henning:2018kys,Maltoni:2019aot}. The underlying reason is as following: according to Goldstone equivalence theorem (GET)~\cite{Cornwall:1974km,Chanowitz:1985hj,Gounaris:1986cr}, scattering amplitudes of longitudinal vector bosons can be approximately evaluated by amplitudes of the corresponding Goldstone bosons i.e. $V_L\sim \phi$.  Moreover, since Goldstone bosons and the Higgs boson form a $SU(2)$ doublet in the SM, as well as in the Standard Model effective field theory (SMEFT)~\cite{Weinberg:1979sa,Buchmuller:1985jz,Leung:1984ni}, couplings of the Goldstone bosons are related to Higgs couplings through the same parameters. Therefore, processes involving longitudinal vector bosons provide an alternative approach to measure Higgs couplings.

Recently, it was proposed in~\cite{Henning:2018kys} that  vector boson scattering (VBS) processes with multiple final states at the LHC with or even without Higgs involved (and its counterparts at lepton colliders) can be used for the measurement of trilinear Higgs coupling.  It was argued that the energy increase of dim-6 operators from longitudinal vector boson enhances the sensitivity of amplitudes to related Wilson coefficients in high energy. As a result,  those process can potentially  be very beneficial to the precise measurement of Higgs self-couplings.  In this work we  follow up this proposal by studying $2\rightarrow 3$ VBS processes extensively at different colliders.  Different from~\cite{Henning:2018kys}, however,  our strategy is to take GET directly and analyze the high energy behavior of  $2\rightarrow 3$ VBS amplitudes under SMEFT.  The goal of our paper is two folded. First, we try to understand more clearly how higher dimension operators affect the amplitudes of $2\rightarrow 3$ VBS. This is mainly achieved by  analyzing how different Feynman diagrams (after taking GET) contribute to the  amplitude. We choose $W_LW_L\rightarrow W_LW_Lh$ and $W^+_LW^-_L\rightarrow hhh$ as examples with  $V_LV_Lh$ and  $hhh$ final states respectively.  Second,  guided by the results of analyzing amplitudes, we carry out simulations to study the measurement of Higgs self-couplings  at the HL-LHC, as well as its future upgrades (HE-LHC), 100 TeV pp colliders and lepton colliders~\cite{CEPCStudyGroup:2018ghi, An:2018dwb,Tian:2013yda, Asner:2013psa,Bambade:2019fyw,Battaglia:2004mw,Robson:2018zje,Roloff:2019crr,Delahaye:2019omf,Long:2020wfp,Buttazzo:2018qqp,Costantini:2020stv,Han:2020pif,Arkani-Hamed:2015vfh}\footnote{Some previous studies for exploring trilinear Higgs coupling can be found in~\cite{He:2015spf,Azatov:2015oxa,deFlorian:2017qfk,Liu-Sheng:2017pxk,Jana:2017hqg,Kim:2018uty,Maltoni:2018ttu,Bizon:2018syu,Biekotter:2018jzu,Borowka:2018pxx,Li:2019uyy,Capozi:2019xsi,Chiesa:2020awd}.}. We set to give a qualitative picture of  the sensitivity of full processes at colliders to dim-6 operators, and the potential of the measurement of Higgs self-couplings at different colliders.

Our main results are briefly summarized as following. At high energy regions, amplitudes of $2\rightarrow 3$ VBS are indeed sensitive to dim-6 operators, with $\frac{\mathcal{A}^{\text{BSM}}}{\mathcal{A}^{\text{SM}}} \sim \frac{E^2}{\Lambda^2}$. However, there are some subtleties involved that will be discussed carefully in the paper.  This sensitivity to dim-6 operators translates further to full processes at colliders, although smallness of cross sections indicates that the processes can only be useful at the future 100 TeV pp colliders or high energy muon colliders.   After exclusively  selecting  longitudinal polarizations  for vector bosons in the final states and applying suitable  $p_T$ cuts in the phase space of final state particles, processes with final state $WWh$ are found to be as important as triple Higgs production ($hhh$).

The rest of the paper is organized as following. In~\autoref{sec:SMEFT}, we lay down the framework of SMEFT and discuss the dim-6 operators that are relevant in this paper and then derive and discuss related scalar couplings.
Then, we derive and analyze the amplitudes of $W_LW_L\rightarrow W_LW_Lh$ and $W^+_LW^-_L\rightarrow hhh$ at high energy by using GET in~\autoref{sec:subprocess}. The dependence on Wilson coefficients are also discussed, along with other subtleties.
The cross section for the full processes $pp\rightarrow jj W_LW_Lh$ and $pp\rightarrow jjhhh$ at hadron colliders and $\mu^+ \mu^-\rightarrow\nu_{\mu}\overline{\nu}_{\mu} W^+_LW^-_Lh$ and $\mu^+ \mu^-\rightarrow\nu_{\mu}\overline{\nu}_{\mu}hhh$ at muon colliders are studied in~\autoref{sec:simulation} through which we discuss the sensitivity of these channels on the SMEFT operators. Finally, we conclude in~\autoref{sec:conclusion}.

\section{Scalar Couplings in SMEFT}

\label{sec:SMEFT}

\subsection{Relevant Dim-6 Operators in SMEFT}
\label{sec:dim_6}

The null result of searching BSM signals at the LHC indicates that new physics may be hidden at the energy scale much higher than the electroweak (EW) scale.  This justifies the usage of effective field theory (EFT) to constrain the possible new physics in a model independent way.  Preserving the SM gauge symmetry group of $SU(3)_c\times SU(2)_L\times U(1)_Y$ further reduces the framework to SMEFT. This framework is suitable for the scenario that the Higgs boson is an elementary particle~\cite{Agrawal:2019bpm}. We will focus on this case and ignore other exotic scenarios hereafter.

Generally, if we ignore the dim-5 Weinberg operator~\cite{Weinberg:1979sa}, the Lagrangian for SMEFT can be written as
\begin{equation}\label{eq:lagrangian}
{\mathcal L}={\mathcal L}_{\text{SM}}+\sum_i\frac{c_i\mathcal{O}_i}{\Lambda^2}+ {\mathcal O}(\frac{1}{\Lambda^3})
\end{equation}
The first term ${\mathcal L}_{\text{SM}}$ is the Lagrangian for the SM, which includes all known physics; the second term includes all dim-6 operators that are suppressed by $\Lambda^2$, with $\Lambda$ being the energy scale of new physics and $c_i$ being Wilson coefficients of the corresponding operators $\mathcal{O}_i$.

Ignoring CP violating terms,  dim-6 operators  relevant to couplings  of (and between) scalars and gauge bosons can be written as
\begin{eqnarray}
\label{eq:EFT_operator}
{\mathcal L}_{\text{dim}-6}=& \frac{1}{\Lambda^2}\left( c_6 (\Phi^\dag\Phi)^3 + c_{\Phi_1} \partial^{\mu}(\Phi^\dag\Phi)\partial_{\mu}(\Phi^\dag\Phi) + c_{\Phi_2} (\Phi^\dag D^{\mu}\Phi)^*(\Phi^\dag  D_{\mu}\Phi)\right. \nonumber \\
                    & + c_{\Phi^2W^2}\Phi^\dag\Phi W_{\mu\nu}^aW^{a\mu\nu}  +  c_{\Phi^2B^2} \Phi^\dag\Phi B_{\mu\nu}B^{\mu\nu}+ c_{\Phi^2WB}\Phi^\dag\tau^a\Phi W_{\mu\nu}^aB^{\mu\nu}\nonumber \\
                    &\left.+ c_{W^3}\epsilon^{abc}W_{\mu}^{a\nu}W_{\nu}^{b\rho}W_{\rho}^{b\mu}\right)
\end{eqnarray}

where $\Phi$ is the Higgs doublet and can be parameterized as
\begin{align}
\Phi=\left(\begin{array}{c}
    \phi^+ \\
    \frac{v+h+i\phi^0}{\sqrt{2}}
    \end{array}\right)
\end{align}

The processes considered in this paper are $2\rightarrow 3$ VBS with initial and final vector bosons being longitudinal polarized, $V_LV_L\rightarrow V_LV_Lh $ and $V_LV_L\rightarrow hhh$. In general, after EWSB, $\phi^\pm/\phi^0$ are considered as unphysical, ``eaten" by gauge fields $W^{\pm}/Z$ becoming their longitudinal components.  The ``real" identities of those degrees of freedom only reveal themselves at high energy through GET~\cite{ Chanowitz:1985hj,Cornwall:1974km,Bagger:1989fc,Gounaris:1986cr,Veltman:1989ud,Yao:1988aj,He:1992nga,He:1993yd,He:1994br}.  Especially in unitary gauge, $\Phi$ becomes $\Phi=\left(0, (v+h)/\sqrt{2}\right)^T$ under which the would-be Goldstone bosons disappear from the physical spectrum and Feynman diagrams.     The three operators in the first line of~\autoref{eq:EFT_operator}, $\mathcal{O}_6$, $\mathcal{O}_{\Phi_1}$ and $\mathcal{O}_{\Phi_2}$,  induce and modify Higgs self-couplings, as well as couplings between vector boson bosons and the Higgs boson.   The operators in the second line and the third line in Eq.(\ref{eq:EFT_operator}) are also involved in scatterings of longitudinal vector bosons, because in this physical picture, couplings involving longitudinal vector bosons, especially vector bosons couplings  to vector bosons,  are mainly induced  by   gauge fields.
However, there is a well-known problem in this physical picture of only gauge bosons being physical: power counting becomes invalid.  This becomes a major obstacle to understanding the leading energy behavior of processes involving longitudinal vector bosons.  Especially, in the presence of SMEFT, higher dimensional operators usually have derivative couplings that lead to unitarity violating energy increase of S-matrix which  are suppressed by scale $\Lambda$. Meanwhile, longitudinal polarization vectors also bring energy increase in the diagram by diagram level, which nevertheless disappears in the final S-matrix as guaranteed by GET.  The real energy increase from derivative couplings and spurious energy increase from longitudinal vector bosons mix together in unitary gauge and the related physical picture. This obscures the underlying physics greatly.

To solve the problem discussed above, we simply take GET and identify $V_L$ with $\phi$ directly.  This is a good approximation as long as the energy scale is much larger than EW scale. Without the spurious energy increase from longitudinal polarization vectors, the energy behavior of Feynman diagrams becomes physical.  Thus we can  analyze the amplitudes at the level of single diagrams and obtain the leading energy behavior of the processes.  Moreover, since $\phi^{\pm}/\phi^0$ and $h$ all belong to the same  $SU(2)$ Higgs doublet, the couplings of (and between) Goldstone bosons and Higgs boson, are determined by the same parameters of the Higgs potential. Thus it becomes manifest that we can measure Higgs self-couplings through processes involving $V_L$s or $\phi$s.

Now let's review the dim-6 operators in \autoref{eq:EFT_operator} under GET.  Higgs self-couplings and couplings between Goldstone bosons and Higgs are induced by  ${\mathcal O}_6$, ${\mathcal O}_{\Phi_1}$ and ${\mathcal O}_{\Phi_2}$, together with SM Lagrangian terms.   ${\mathcal O}_6$  is the only one that contributes to  5-point and 6-point scalar vertices.  ${\mathcal O}_{\Phi_2}$ term violates custodial symmetry, the Wilson coefficient $c_{\Phi_2}$ is strongly constrained by LEP~\cite{Azatov:2015oxa}. Therefore, we will ignore it from now on. ${\mathcal O}_{\Phi_1}$, ${\mathcal O}_{\Phi^2W^2}$, ${\mathcal O}_{\Phi^2B^2}$ and  ${\mathcal O}_{\Phi^2WB}$ give rise to $\text{gauge-gauge-scalar}$ and $\text{gauge-gauge-scalar-scalar}$ vertices. ${\mathcal O}_{W^3}$ doesn't contribute to $V_LV_L\rightarrow V_LV_Lh$, but can have contributions to the amplitudes if the polarizations of vector bosons are transverse. Moreover, ${\mathcal O}_{W^3}$'s vertices also contribute to $2\to4$ process, e.g. $V_LV_L\rightarrow V_LV_Lhh$. For simplicity, we are only interested in the modification to scalar couplings in this work. Thus only $\mathcal{O}_6$ and $\mathcal{O}_{\Phi_1}$ are considered.

\subsection{Scalar Vertices}
\label{sec:mass_basis}

The full Feynman rules in SMEFT can be found in~\cite{Dedes:2017zog}. Here we briefly review the results related to our processes.
Before symmetry breaking, the Lagrangian in scalar sector is
\begin{equation}
{\mathcal L}_{\Phi}=(D_\mu\Phi)^\dag (D^\mu\Phi)  +\mu^2\Phi^\dag\Phi-\lambda_h(\Phi^\dag\Phi)^2+\frac{c_6}{\Lambda^2}(\Phi^\dag\Phi)^3+\frac{c_{\Phi_1}}{\Lambda^2} \partial^{\mu}(\Phi^\dag\Phi)\partial_{\mu}(\Phi^\dag\Phi),
\end{equation}
where $\Lambda$ is the new physics scale,   which we  choose as  $1$ TeV in the paper.  After symmetry breaking, the VEV $v$ can be expressed in terms of $\mu$, $\lambda_h$ and $c_6$ by tracking the minimum position of the Higgs potential,
\begin{align}
\label{eq:vev}
v=\sqrt{\frac{\mu^2}{\lambda_h}}- \frac{3}{8}\frac{\mu^3}{\lambda_h^{5/2}}\frac{c_6}{\Lambda^2}
\end{align}
The Higgs field $h$ and goldstone field $\phi^0$ need extra field renormalization, after which, we obtain Higgs mass $m_h$ in terms of $(v, c_6, c_{\Phi_1}, \lambda_h)$:
\begin{align}
\label{eq:mh2}
m_h^2=\frac{1}{2}\lambda_hv^2 -(3c_6 + \lambda_h c_{\Phi_1})\frac{v^4}{\Lambda^2}
\end{align}
Eliminating $(\mu, \lambda_h)$ by $(m_h, v)$ through~\autoref{eq:vev} and \autoref{eq:mh2}, we can express all scalar vertices in terms of  $(m_h, v, c_6, c_{\Phi_1})$, with $v^2=\frac{\sqrt{2}g}{2G_F}\approx(246\,{\rm GeV})^2$.
Then, we have  all 3-point, 4-point and 5-point scalar couplings to be
\begin{subequations}
\begin{align}
\lambda_{h \phi^+\phi^-} &= \lambda_{h\phi^0\phi^0} = -i\frac{m_h^2}{v} +i c_{\Phi_1}v\frac{2p_h^2+m_h^2}{\Lambda^2}, \\
\lambda_{hhh} &= -i\left(\frac{3m_h^2}{v}  - 6c_6\frac{v^3}{\Lambda^2}\right)  + ic_{\Phi_1}v\frac{2p_1^2+2p_2^2+2p_3^2+3m_h^2}{\Lambda^2}, \\
\lambda_{\phi_0^4} &= -3i\frac{m_h^2}{v^2} -3ic_{\Phi_1}\frac{v^2}{\Lambda^2}-2i c_{\Phi_1}\frac{(p_1+p_2)^2+(p_1+p_3)^2+(p_1+p_4)^2}{\Lambda^2}, \\
\lambda_{\phi_+\phi_-\phi_0^2} &= -i\frac{m_h^2}{v^2} -ic_{\Phi_1}\frac{v^2}{\Lambda^2}-2ic_{\Phi_1}\frac{(p_1+p_2)^2}{\Lambda^2}, \\
\lambda_{\phi_+^2\phi_-^2} &=-2i\frac{m_h^2}{v^2} -2ic_{\Phi_1}\frac{v^2}{\Lambda^2} \nonumber \\
&\qquad -2ic_{\Phi_1}\frac{2(p_1+p_3)^2+ 2(p_1+p_4)^2-p_1^2-p_2^2-p_3^2-p_4^2}{\Lambda^2},  \\
\lambda_{\phi_+\phi_-h^2} &=\lambda_{\phi_0^2h^2}=-i\left(\frac{m_h^2}{v^2}-6c_6\frac{v^2}{\Lambda^2}+c_{\Phi_1}\frac{v^2}{\Lambda^2}\right)-2ic_{\Phi_1}\frac{(p_1+p_2)^2+m_h^2}{\Lambda^2},\\
\lambda_{h^4}&= -3i\left(\frac{m_h^2}{v^2}-12c_6\frac{v^2}{\Lambda^2}+c_{\Phi_1}\frac{v^2}{\Lambda^2}\right) \nonumber \\
&\qquad -2ic_{\Phi_1} \frac{(p_1+p_2)^2+(p_1+p_3)^2+(p_1+p_4)^2+6m_h^2}{\Lambda^2},\\
\lambda_{\phi^+\phi^-h^3}&=\lambda_{(\phi^0)^2h^3}=\lambda_{(\phi^0)^4h}=18ic_6\frac{v}{\Lambda^2},  \\
\lambda_{(\phi^+\phi^-)^2h}&= 12ic_6 \frac{v}{\Lambda^2},\\
\lambda_{\phi^+\phi^-(\phi^0)^2h}&=6ic_6 \frac{v}{\Lambda^2},
\end{align}
\end{subequations}
where $p_h$ in $\lambda_{h \phi^+\phi^-}$ and  $\lambda_{h\phi^0\phi^0}$ is the momentum of the Higgs boson; $p_1, p_2$ in $\lambda_{\phi_+\phi_-\phi_0^2}$, $\lambda_{\phi_+\phi_-h^2}$ and $\lambda_{\phi_0^2h^2}$ are momenta of $\phi^+/\phi^-$ or equivalently $h$; in $\lambda_{\phi_+^2\phi_-^2}$, we can assign $p_1$ and $p_2$ to $\phi^+$, $p_3$ and $p_4$ to $\phi^-$. 6-point scalar couplings are not listed as they are irrelevant to the processes we study here.

\section{Amplitudes and Cross Sections}
\label{sec:subprocess}

\subsection{Feynman Diagrams and Amplitudes with Goldstone Equivalence}

\label{sec:sub_amp}

In this section we analyze the amplitude of $V_LV_L\rightarrow V_LV_L h$ and $V_LV_L\rightarrow hh h$ at high energy by using GET for SM plus two dim-6 operators of SMEFT ($\mathcal{O}_6$, $\mathcal{O}_{\Phi_1}$). For simplicity, we will choose $W^+_LW^-_L\rightarrow W^+_LW^-_L h$ and $W^+_LW^-_L\rightarrow hh h$ as examples to illustrate the general behavior, and work in Feynman gauge.

Following GET, $W^+_LW^-_L\rightarrow W^+_LW^-_L h$ and $W^+_LW^-_L\rightarrow hh h$ can be approximated by  $\phi^+(p_1)\phi^-(p_2)\rightarrow \phi^+(p_3)\phi^-(p_4) h(p_5)$ and $\phi^+(p_1)\phi^-(p_2)\rightarrow h(p_3) h(p_4) h(p_5)$ at high energy. The corresponding amplitudes can be classified according to the number of propagators: zero, one or two propagators:
\begin{equation}
{\mathcal A}= {\mathcal A}_0+{\mathcal A}_1+{\mathcal A}_2,
\end{equation}
where $\mathcal{A}_i$ with $i=0,1,2$ can be further classified as diagrams from SM vertices only and diagrams including BSM contributions: $\mathcal{A}_i=\mathcal{A}_i^{\text{SM}}+\mathcal{A}_i^{\text{BSM}}$.

Before going to the analysis in details, it's important to understand in a general way how the high energy behavior of $\phi\phi\rightarrow \phi\phi h$ is determined. After the unphysical energy increase from longitudinal polarization vectors getting eliminated,  physical energy dependence only comes from derivative couplings in  dim-6 operators as well as the $\frac{1}{E^2}$ factor of the propagators.  Understanding the overall energy behavior of the amplitude from the interplay of different factors is the main focus of our analysis. We mainly focus on  $\phi^+\phi^-\rightarrow \phi^+\phi^- h$ process, and make comments when there is difference with $\phi^+\phi^-\rightarrow h h  h$.    Since we will focus on the high energy behavior,  we also only keep the leading terms in $\frac{1}{E}$.   For ${\mathcal A}^{\text BSM}$, we only keep the leading terms of $\frac{c_i}{\Lambda^2}$. Higher order terms  of $\frac{1}{\Lambda^2}$ are  neglected as they are suppressed by additional powers of $\Lambda$.  Moreover, to be fully consistent when considering higher order terms, we would have to take into account the higher dimensional operators, thus go beyond dim-6 SMEFT.

\begin{figure}
\centering
\includegraphics[width=0.3 \textwidth]{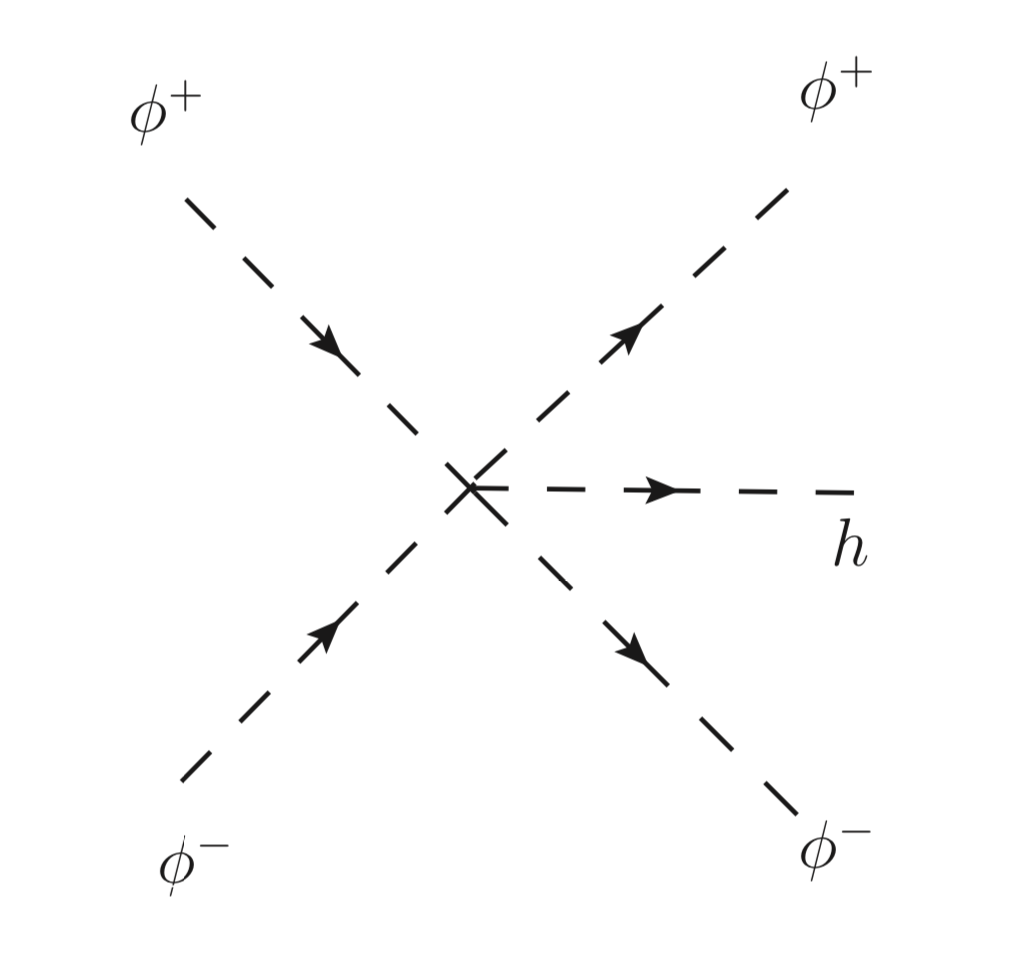}
\caption{Contact diagram for $\phi^+\phi^-\rightarrow \phi^+\phi^-h$}
\label{fig:contact}
\end{figure}

\begin{figure}
\centering
\includegraphics[width=0.3 \textwidth]{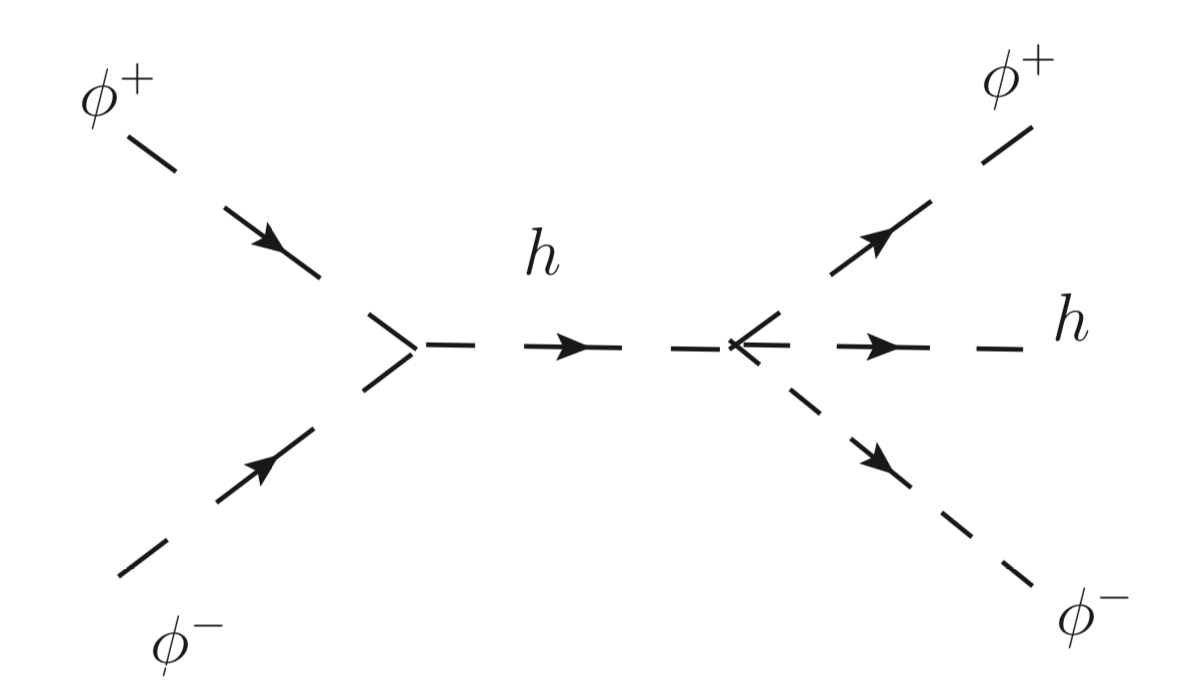}
\caption{A typical diagram with one propagator for $\phi^+\phi^-\rightarrow \phi^+\phi^-h$ }
\label{fig:one-prop}
\end{figure}

\subsubsection*{Diagrams with $0$ propagator}

In the SM, $\phi^+\phi^-\rightarrow \phi^+\phi^- h$ and $\phi^+\phi^-\rightarrow h h  h$ have no 5-point contact diagram. There is, however, one such diagram from the $\mathcal{O}_6$ operator, as shown in~\autoref{fig:contact}. The corresponding amplitudes are
\begin{align}
    \label{eq:amp_c6}
    \mathcal{A}_0^{\phi^+\phi^-\rightarrow \phi^+\phi^- h}&= \lambda_{(\phi^+\phi^-)^2h}= 12ic_6\frac{v}{\Lambda^2}\\
    \label{eq:amp_c6_2}
    \mathcal{A}_0^{\phi^+\phi^-\rightarrow hh h}&= \lambda_{\phi^+\phi^-h^3}= 18ic_6\frac{v}{\Lambda^2}
\end{align}
In both cases we obtain $\mathcal{A}_0\sim \frac{v}{\Lambda^2}$.

\subsubsection*{Diagrams with $1$ propagator}

For any $2\to3$ process, the amplitude of all Feynman diagrams with one propagator can be schematically written as
\begin{equation}
\mathcal{A}_1 =\sum_{\text{diagrams}}i\frac{  M'_4\cdot  M'_3}{q^2-m^2}
\end{equation}
The intermediate particle of the propagator can be either a scalar or a vector boson, as illustrated in~\autoref{fig:one-prop} for $\phi^+\phi^-\to\phi^+\phi^-h$. $M'_4$ and $M_3'$ denote the amplitudes involving 4-point and 3-point vertices respectively, with Lorentz indices suppressed if the intermediate state is a vector boson.  However, for the vector boson propagator, we need $\phi\phi\phi V$-like vertex, which does not exist in the SM and dim-6 operators we consider in this work ($\mathcal{O}_6$ and $\mathcal{O}_{\Phi_1}$)\footnote{We could have $\phi\phi\phi V$-like vertex, once we include more operators in SMEFT, e.g. $\mathcal{O}_{\Phi_2}$. The analysis will be similar and they have same behavior at high energy.}. Thus the propagator can only be a scalar.

For SM-only diagrams, we have
\begin{align}
 \mathcal{A}_1^{\text{SM}} =& -2i\frac{m_h^2}{v^3} \left(\frac{1}{(p_3+p_5)^2-m_W^2} + \frac{1}{(p_4+p_5)^2-m_W^2} + \frac{1}{(p_2-p_5)^2-m_W^2} + \frac{1}{(p_1-p_5)^2-m_W^2}\right) \nonumber \\
               & -2i\frac{m_h^2}{v^3}\left(\frac{1}{(p_1+p_2)^2-m_h^2}+ \frac{1}{(p_1-p_3)^2-m_h^2}\right).
\end{align}
Neither 4-point vertices nor 3-point vertices have any energy dependence. Since at high energy, $p_i^2 \sim E^2 \gg m_W^2, m_h^2$, the amplitude scales as  $\mathcal{A}_1^{SM}\sim \frac{v}{E^2}$.

For BSM contributions, keep terms up to $\frac{1}{\Lambda^2}$, we have:
\begin{align}
\mathcal{A}_1^{BSM} \simeq & -i 2c_{\Phi_1}\frac{m_h^2}{v\Lambda^2}\left(\frac{(p_1+p_2)^2}{(p_4+p_5)^2-m_W^2} +\frac{(p_1+p_2)^2}{(p_3+p_5)^2-m_W^2}  + \frac{(p_1-p_3)^2}{(p_2-p_5)^2-m_W^2} +\frac{(p_2-p_4)^2}{(p_1-p_5)^2-m_W^2}   \right)   \nonumber \\
                &- i c_{\Phi_1}\frac{m_h^2}{v\Lambda^2} \left(\frac{(p_1+p_2)^2}{(p_3+p_4)^2-m_h^2} +\frac{(p_3+p_4)^2}{(p_1+p_2)^2-m_h^2}  + \frac{(p_1-p_3)^2}{(p_2-p_4)^2-m_h^2} +\frac{(p_2-p_4)^2}{(p_1-p_3)^2-m_h^2}\right)\nonumber \\
                &-16ic_{\Phi_1}\frac{m_h^2}{v\Lambda^2}\left(\frac{(p_3+p_4)^2}{(p_3+p_4)^2-m_h^2} +\frac{(p_1+p_2)^2}{(p_1+p_2)^2-m_h^2}  + \frac{(p_2-p_4)^2}{(p_2-p_4)^2-m_h^2} +\frac{(p_1-p_3)^2}{(p_1-p_3)^2-m_h^2}\right),
\end{align}
where both 3-point and 4-point scalar vertices provide momentum-dependent couplings leading to $E^2$ behavior in the numerator, which cancels the $\frac{1}{E^2}$ factor from the propagator. Therefore, we obtain $\mathcal{A}_1^{BSM}\sim \frac{v}{\Lambda^2} $. Similarly, the behavior of amplitude for $\phi^+\phi^-\rightarrow hhh$ also scales as $\frac{v}{E^2}$ in the SM and $\frac{v}{\Lambda^2}$ when dim-6 operators are involved.

\subsubsection*{Diagrams with 2 propagators}

\begin{figure*}[t!]
\centering
\begin{subfigure}[t]{0.3\textwidth}
\includegraphics[width=\textwidth]{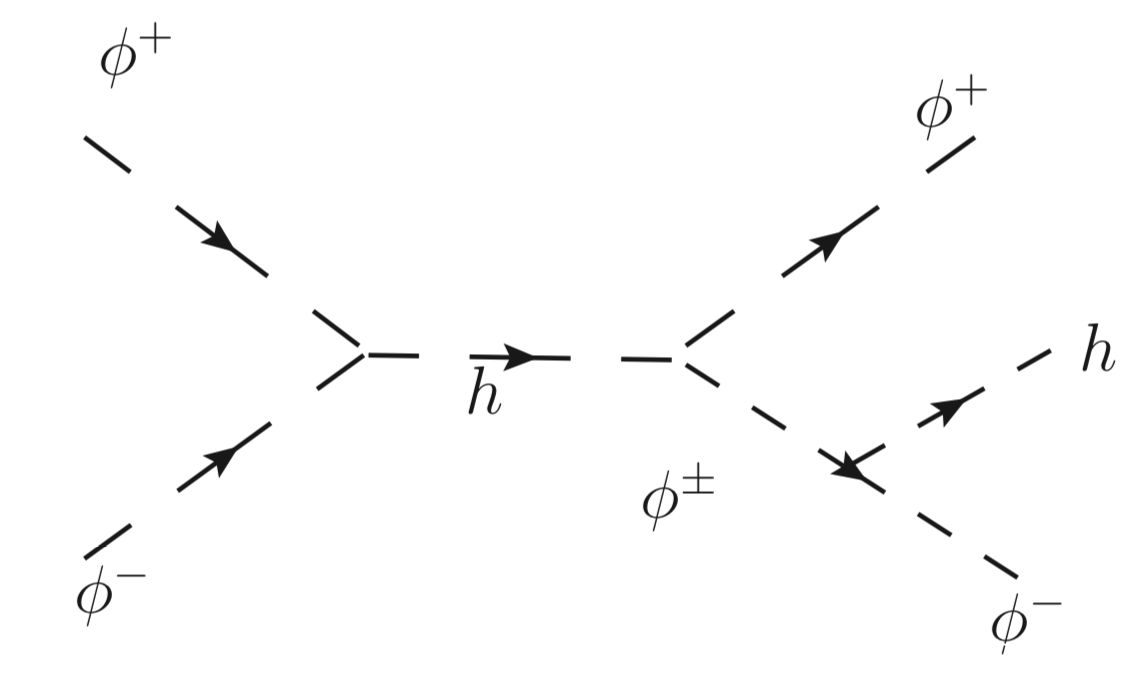}
\caption{Two scalar propagators}
\label{fig:two-prop-1}
\end{subfigure}
\begin{subfigure}[t]{0.3\textwidth}
\includegraphics[width=\textwidth]{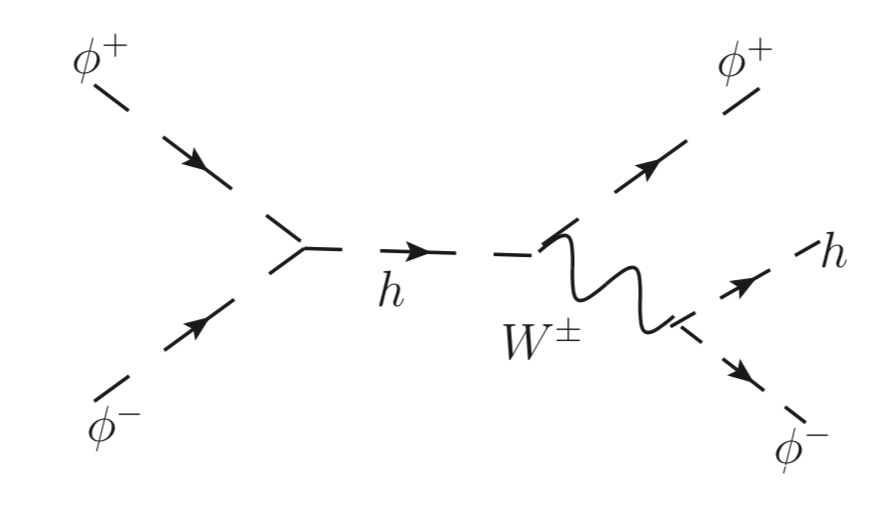}
\caption{One scalar propagator, one gauge propagator}
\label{fig:two-prop-2}
\end{subfigure}
\begin{subfigure}[t]{0.3\textwidth}
\includegraphics[width=\textwidth]{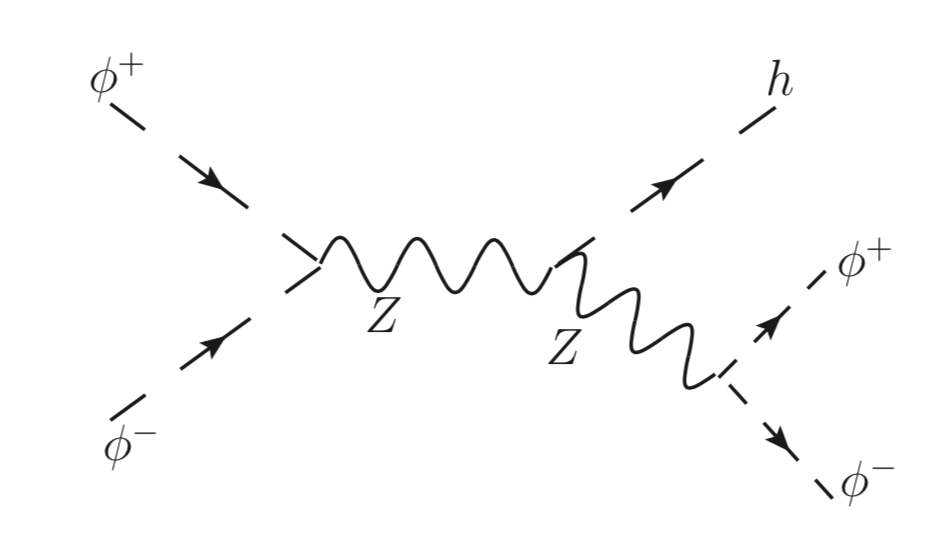}
\caption{Two gauge propagators}
\label{fig:two-prop-3}
\end{subfigure}
\caption{Typical Diagrams with two propagators for $\phi^+\phi^-\rightarrow \phi^+\phi^-h$}
\label{fig:two-prop}
\end{figure*}

The amplitudes for the Feynman diagrams with two propagators can be classified according to the type of the propagators as shown in~\autoref{fig:two-prop}:
\begin{description}
    \item[(a)] Two scalars
    \item[(b)] One scalar and one vector boson
    \item[(c)] Two vector bosons
\end{description}
There are too many diagrams with two propagators to give a short and concise analytical expression even in high energy limit, hence we only analyze a representative diagram for each case.

For the case with two scalar propagators, a typical diagram is shown in~\autoref{fig:two-prop-1} which has both SM and BSM contributions:

\begin{align}
\mathcal{A}_2^{a,\text{SM}}\simeq& -i(\frac{m_h^2}{v})^3\frac{1}{(p_1+p_2)^2-m_h^2} \frac{1}{(p_4+p_5)^2-m_W^2},  \nonumber \\
\mathcal{A}_2^{a,\text{BSM}}\simeq& 4ic_{\Phi_1}\frac{m_h^4}{v\Lambda^2}\frac{(p_1+p_2)^2}{(p_1+p_2)^2-m_h^2}\frac{1}{(p_4+p_5)^2-m_W^2}.
\end{align}
Thus, at high energy, diagrams with two scalar propagators scale as $\mathcal{A}_2^{a,\text{SM}}\sim \frac{v^3}{E^4}$, $\mathcal{A}_2^{a,\text{BSM}}\sim \frac{v^3}{\Lambda^2E^2}$. They are suppressed by $\frac{1}{E^2}$ compared with $\mathcal{A}_1^{\text{SM}}$ and $\mathcal{A}_1^{\text{BSM}}$ respectively.

The typical diagram for the case with one scalar and one vector boson propagators is shown in~\autoref{fig:two-prop-2}. Summing the diagram and the diagram by exchanging $p_3\leftrightarrow p_4$, keeping only the leading contributions, we have
\begin{align}
\mathcal{A}_2^{b,\rm SM}\simeq &i\frac{g^2m_h^2}{4v}
\left(\frac{(p_1+p_2+p_3)(p_5-p_4)}{((p_1+p_2)^2-m_h^2)((p_4+p_5)^2-m_W^2)} + \frac{(p_1+p_2+p_4)(p_5-p_3)}{((p_1+p_2)^2-m_h^2)((p_3+p_5)^2-m_W^2)} \right),
\end{align}
where we only have SM contribution. At high energy, the amplitude scales as $\mathcal{A}_2^{b,SM}\sim \frac{v}{E^2}$, which is of the same order as ${\mathcal A}_1^{\text{SM}}$.

In~\autoref{fig:two-prop-3}, we show the typical diagram for the case with two vector boson propagators. The amplitude only receives SM contributions and reads:
\begin{align}
\mathcal{A}_2^c\simeq& =-i\frac{g^3m_Zc_{2W}^2}{4c_W^3} \frac{(p_1-p_2)\cdot (p_3- p_4)}{((p_1+p_2)^2-m_Z^2)((p_3+p_4)^2-m_Z^2)},
\end{align}
where $c_{2W}=\cos2\theta_W$, $c_W=\cos\theta_W$ and $\theta_W$ is the Weinberg angle.
At high energy, it scales as $\mathcal{A}^c_2\sim \frac{v}{E^2}$.

\subsubsection*{Combined amplitudes}

Taking into account all cases discussed above, the amplitude of $W_L^+W_L^-\to W_L^+W_L^-h$ can be written as

\begin{align}
\mathcal{A}(W_L^+W_L^-\rightarrow W_L^+W^-_Lh)=\mathcal{A}^{\text{SM}} + \mathcal{A}^{\text{BSM}},
\end{align}
with  $\mathcal{A}^{\rm SM}$ being the SM contribution which has no dependence on $c_6$ or $c_{\Phi_1}$ and $\mathcal{A}^{\rm BSM}$ the BSM contributions depending on $c_6$ and $c_{\Phi_1}$.  We only keep the terms up to the order of $\frac{c_i}{\Lambda^2}$,  higher order terms are truncated to be consistent with the EFT expansions.

The leading energy dependence of SM and BSM contributions are
\begin{align}
\label{eq:ASM_ABSM}
\mathcal{A}^{\text{SM}} \sim \frac{v}{E^2}, \quad \mathcal{A}^{\text{BSM}} \sim \frac{v}{\Lambda^2}.
\end{align}
Thus, the ratio between BSM and SM is approximately
\begin{align}
\label{eq:amp-ratio}
 \frac{\mathcal{A}^{BSM}}{\mathcal{A}^{SM}}\sim  \frac{E^2}{\Lambda^2}.
\end{align}

From~\autoref{eq:amp-ratio} we find that, the BSM contribution to the total amplitude will be enhanced relative to the SM one at high energy. Let's stop and analyze  the physical reasons behind~\autoref{eq:amp-ratio}. By a naive dimensional analysis, the amplitude for $2\to3$ process will scale as $1/(\rm GeV)$.
For the SM contributions, combining the energy dependence from the propagator and 3-point vertices, it scales as $\frac{v}{E^2}$ with the energy  coming from the propagator.
On the other hand, the BSM contribution has a different leading energy behavior -- it remains constant as ($\frac{v}{\Lambda^2}$) due to: (a) 5-point scalar vertices as shown in~\autoref{fig:contact} from $\mathcal{O}_6$, giving dependence on $c_6$;  (b) The cancellation between energy suppression from propagators and  energy increase from momentum dependence in 3/4-point vertices from $\mathcal{O}_{\Phi_1}$, giving dependence on $c_{\Phi_1}$. Although this enhancement of the BSM contribution from $c_{\Phi_1}$ relative to SM one applies to many processes, it is not the case for $c_6$, which depends crucially on the 5-point scalar vertices coming solely from $\mathcal{O}_6$. Since $\mathcal{O}_6$ is also the only source for  6-point scalar vertices, this sensitivity of amplitude to BSM physics for $c_6$ also applies to $2\rightarrow 4$ VBS processes. In comparison, 4-point scalar vertices can come from both SM and higher dimensional operators, therefore the amplitude of $2\rightarrow 2$ VBS does not have the behavior of~\autoref{eq:ASM_ABSM} for $c_6$.

There is also a subtlety that's related to the so-called soft-collinear singularities that originate from propagators reaching (close to) to the poles, see, for example, \cite{Kinoshita:1962ur,Lee:1964is,Sterman:1977wj,Sterman:1986aj,Dawson:1984gx,Chen:2016wkt}.

After integrating over phase space, those singularities result in logarithmic enhancement to the cross sections which will change the behavior in~\autoref{eq:amp-ratio} and can reduce the sensitivities to the Wilson coefficients. Hence those singularities require careful treatments which we will discuss in the next section.

\subsection{Cross Sections for Subprocesses }

\label{sec:sub_cross}

After deriving the amplitudes of $W_L W_L\rightarrow W_L W_L h$ and $W_L W_L\rightarrow hhh$ using GET, we will now examine the behaviors of the cross sections for $W_L W_L\rightarrow W_L W_L h$ and $W_L W_L\rightarrow hhh$ which is calculated using {\tt FeynArts}~\cite{Hahn:2000kx} and {\tt FormCalc}~\cite{Hahn:1998yk} with a cut $p_T > 50$ GeV on the final states and also cross checked with {\tt MadGraph}~\cite{Alwall:2014hca}. The dependence on $c_6$ and $c_{\Phi_1}$ are considered separately.

\begin{figure}[!tb]
    \centering
    \includegraphics[width=\textwidth]{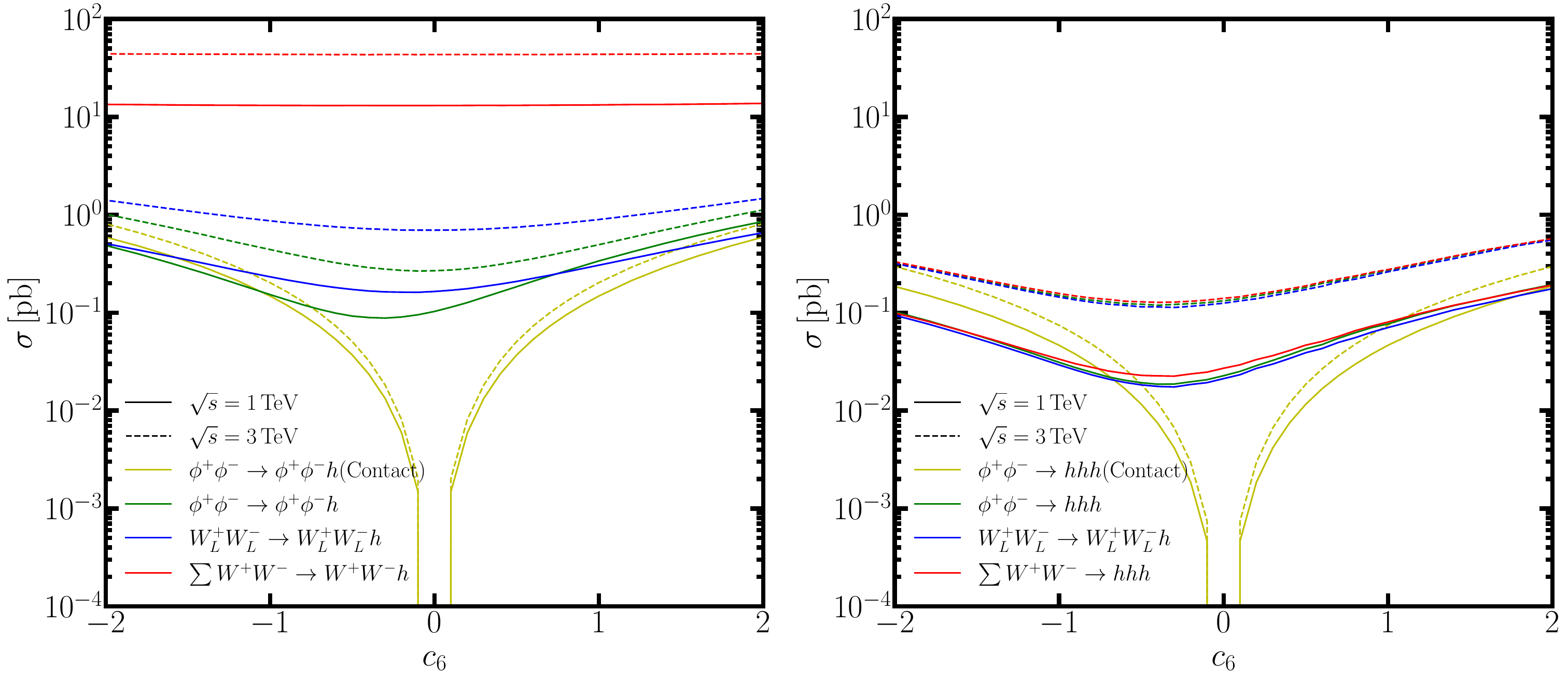}
    \caption{$\hat{\sigma}(W^+_LW^-_L\rightarrow W^+_LW^-_Lh)$ and $\hat{\sigma}(W^+_LW^-_L\rightarrow hhh)$ as   functions of $c_6$. }
    \label{fig:C6-1}
\end{figure}

The dependence of the cross section for $WW\to WWh$ and $WW\to hhh$ on $c_6$ is shown in~\autoref{fig:C6-1} for two representative energy $\sqrt{s}=1\,\rm TeV$ (solid lines) and $\sqrt{s}=3\,\rm TeV$ (dashed lines). The dependence on $c_6$ for these processes only comes from the 5-point contact terms as shown in~\autoref{fig:contact} which is denoted as the yellow line in~\autoref{fig:C6-1}. Thus the dependence of the cross section of $\phi\phi\to\phi\phi h$ and $\phi\phi\to hhh$ (green lines), as well as $W_LW_L\to W_LW_Lh$ and $W_LW_L\to hhh$ (blue lines), mainly follows the behavior of the yellow lines, except the region where $c_6$ is close to zero.

\begin{figure}[!tb]
    \begin{center}
        \includegraphics[width=0.98\textwidth]{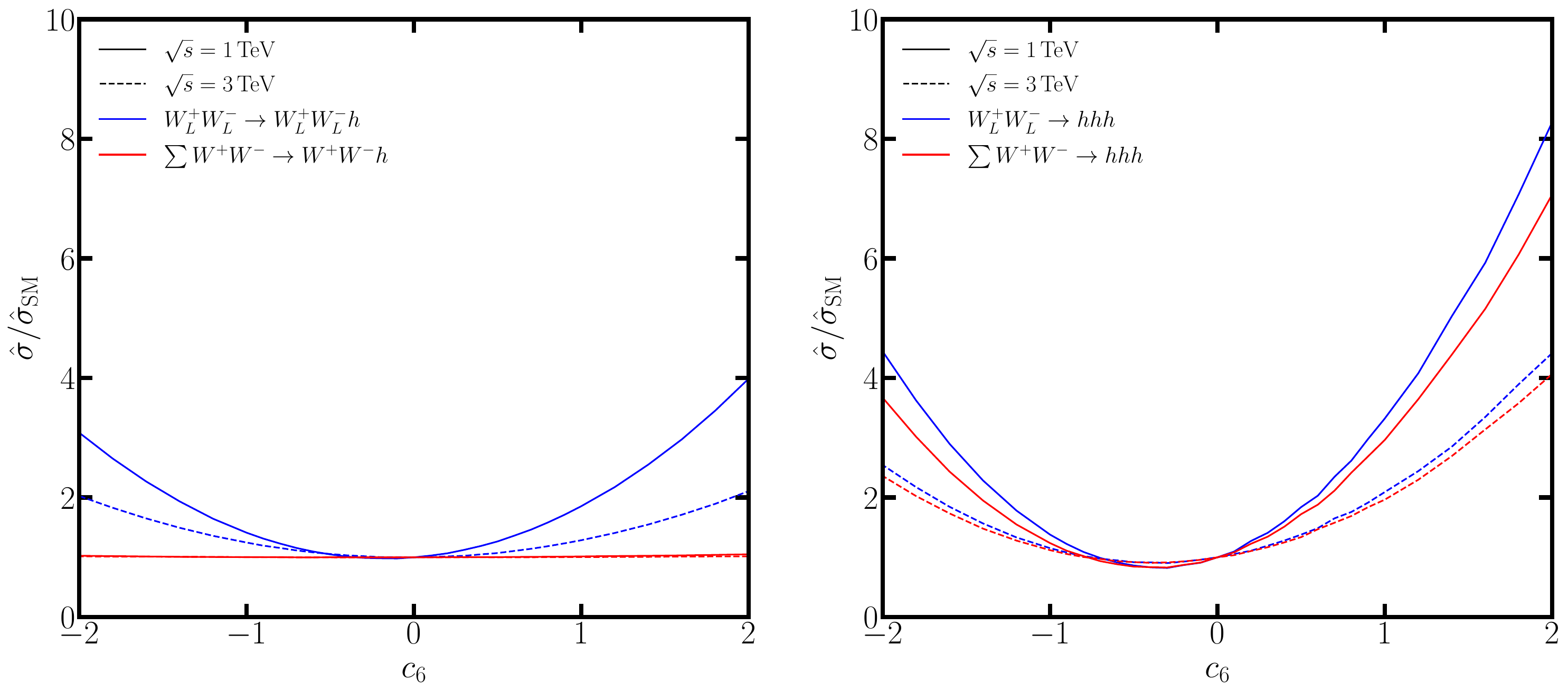}
    \end{center}
    \caption{$\hat{\sigma}/\hat{\sigma}_{\rm SM}$ for $W^+W^-\to W^+W^-h$ and $W^+W^-\to hhh$ as functions of $c_6$.}
    \label{fig:cs_ratio_c6}
\end{figure}

However, $WW\to WWh$ is dominated by transverse polarizations as can be seen by comparing the red and blue lines. Thus, in order for the former process to remain sensitive to $c_6$, longitudinal polarizations of $W$ boson pair need to be singled out by some specific selections. The technical details can be found in~\cite{Ballestrero:2017bxn,Ballestrero:2020qgv,De:2020iwq,Grossi:2020orx,Kim:2021gtv}. On the other hand $WW\to hhh$ is dominated by longitudinal polarizations, hence, the sensitivity is largely remained in this sense. In~\autoref{fig:cs_ratio_c6}, we show the ratio $\sigma/\sigma_{\rm SM}$ as functions of $c_6$ for $W^+W^-\to W^+W^-h$ and $W^+W^-\to hhh$. It is clear that, by just looking at these $2\to3$ processes, $WW\to hhh$ is more sensitive than $WW\to WWh$ on $c_6$. Note that, the sensitivity is reduced at higher energy which is mainly due to the logarithm enhancement of the SM cross section as discussed in previous section.

\begin{figure}[!tb]
    \centering
    \includegraphics[width=\textwidth]{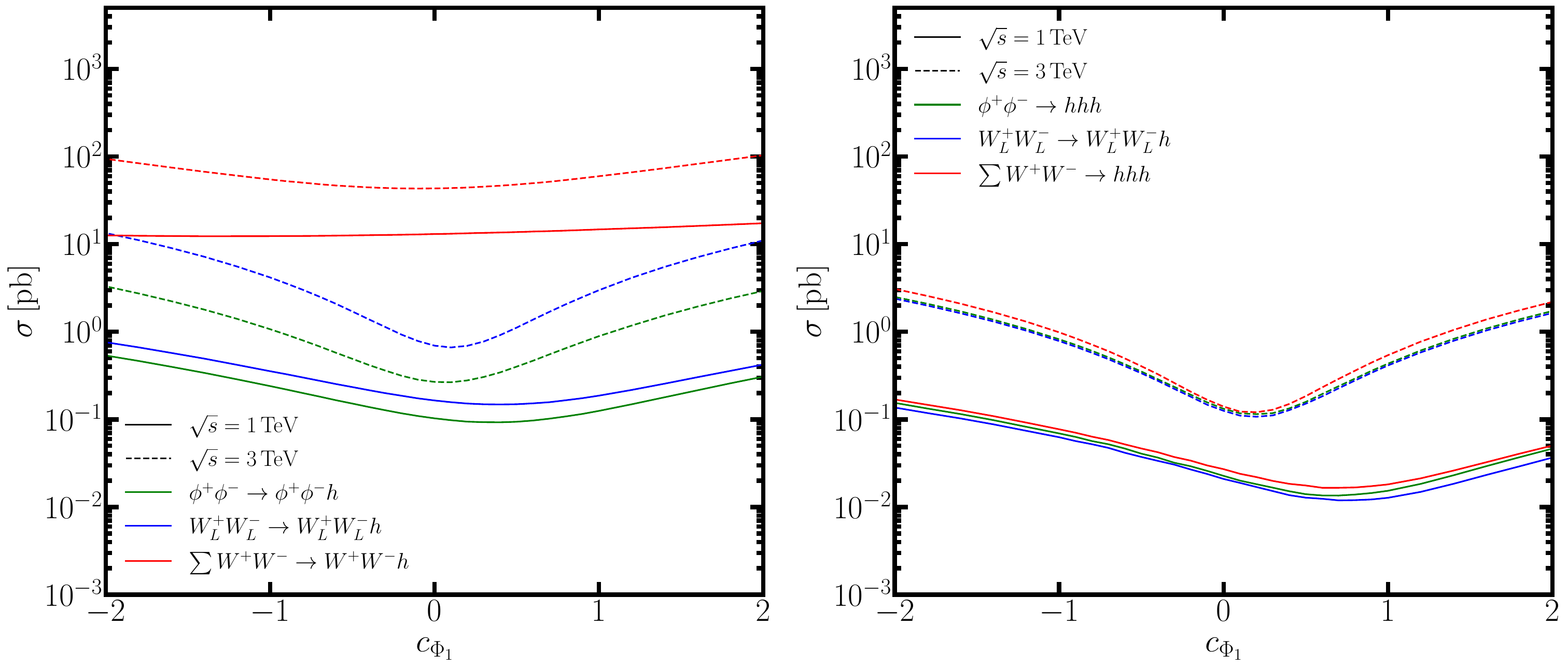}
    \caption{$\hat{\sigma}(W^+_LW^-_L\rightarrow W^+_LW^-_Lh)$ and $\hat{\sigma}(W^+_LW^-_L\rightarrow hhh)$ as   functions of $c_{\Phi_1}$. }
    \label{fig:Cphi-1}
\end{figure}

\begin{figure}[!tb]
    \begin{center}
        \includegraphics[width=\textwidth]{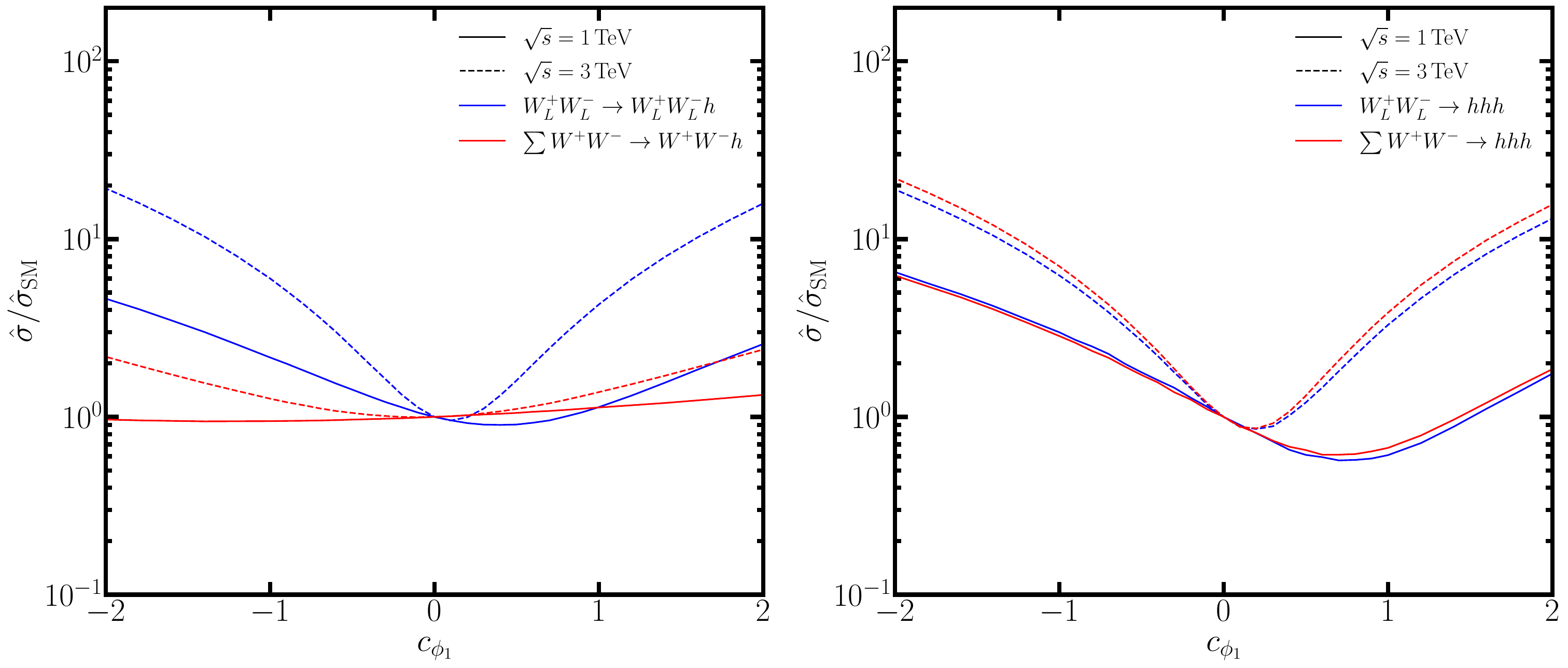}
    \end{center}
    \caption{$\hat{\sigma}/\hat{\sigma}_{\rm SM}$ for $W^+W^-\to W^+W^-h$ and $W^+W^-\to hhh$ as functions of $c_{\Phi_1}$.}
    \label{fig:cs_ratio_cphi}
\end{figure}

Similar results for $c_{\Phi_1}$ are shown in~\autoref{fig:Cphi-1} and~\autoref{fig:cs_ratio_cphi}. Again, $WW\to WWh$ is dominated by transverse polarizations. Compared with the dependence on $c_6$, the cross section is much more enhanced by $c_{\Phi_1}$ due to the momentum dependence of $\mathcal{O}_{\Phi_1}$ operator. Further, this momentum dependence of $\mathcal{O}_{\Phi_1}$ also overcomes the logarithms in SM cross section, thus higher energy corresponds to higher sensitivities.  Finally, we comment that the shapes of  cross sections v.s. $c_6(c_{\Phi_1})$ are parabolic. This is consistent with the fact that only terms up to the order of $\frac{1}{\Lambda^2}$   are kept in the amplitude level. As a result, the cross sections are quadratic functions of $c_6 (c_{\Phi_1})$.

In this section, we discussed the cross section for  $2\to3$ VBS processes and their dependence on $c_6$ and $c_{\Phi_1}$ which provides the basic ideas about the sources of the sensitivities. In the next section, we will proceed to discuss the measurement of $c_6$ and $c_{\Phi_1}$ with more realistic setup at both hadron and lepton colliders.

\section{VBS Processes at Hadron/Lepton Collider}
\label{sec:simulation}

\subsection{Signal Processes}
\label{sec:simulation_dis}

In~\autoref{sec:subprocess}, we analyzed VBS processes with three bosons in the final state at high energy, taking $W_L^+ W_L^-\rightarrow W_L^+ W_L^- h$ and $W_L^+ W_L^-\rightarrow hhh$ as two examples. We found that they are sensitive to dim-6 operators in the SMEFT as shown in~\autoref{fig:cs_ratio_c6} and~\autoref{fig:cs_ratio_cphi}. In this section we continue to analyze the corresponding full processes at hadron and lepton colliders using
{\tt MadGraph}~\cite{Alwall:2014hca} with the SMEFT model file from Ref.~\cite{Degrande:2020evl}.

Since the aim of this paper is to illustrate important and crucial behaviors of $2\rightarrow 3$ VBS processes, we don't plan to cover all related processes here. Instead, we choose the following processes as benchmarks:
\begin{align}
    l^+l^-&\rightarrow\nu_l\bar{\nu}_l W^+_LW^-_L h\qquad    l^+l^-\rightarrow \nu_l \bar{\nu}_l h h h\label{proc:ee}  \\
pp&\rightarrow j j W^{\pm}_LW^{\pm}_L h\qquad\ \ \ \ pp\rightarrow j j h h h \label{proc:pp}
\end{align}
where $l$ is either  $\mu$ or   $e$.
Processes in~\autoref{proc:ee} can be explored at a series of future lepton colliders, including
CLIC~\cite{Battaglia:2004mw,Robson:2018zje,Roloff:2019crr} ($1 \text{TeV}<\sqrt{s}<3 \text{TeV}$) and the recently proposed muon colliders~\cite{Delahaye:2019omf,Long:2020wfp,Buttazzo:2018qqp,Costantini:2020stv,Han:2020pif} ($3 \text{TeV}<\sqrt{s}<30 \text{TeV}$). Processes in~\autoref{proc:pp} can be explored at the HL-LHC, HE-LHC~\cite{Cepeda:2019klc} and the future 100 TeV pp colliders~\cite{Arkani-Hamed:2015vfh,Mangano:2016jyj}.
Notice that we focus on the same-sign $W$s for $pp\rightarrow j j W_L W_L h$ due to the suppressions of relevant SM backgrounds for this process compared with the opposite-sign one.
We will devote a comprehensive survey of all relevant VBS processes in future works.

\subsection{Signal Cross Sections}
\label{sec:simulation_data}

In this section, we will examine the cross section for relevant signal processes listed in last section, especially the dependence of the cross section on $c_6$ and $c_{\Phi_1}$ and the comparison with SM cross section. In the following, we will present the results for $WWh$ and $hhh$ productions respectively.

\subsubsection*{Production of $WWh$}

\begin{table}[!tb]
\centering
\begin{tabular}{ |c|c|c|c|c|c|}
\hline
\multicolumn{6}{|c|}{Cross sections (pb) for $\mu^+ \mu^-\rightarrow \nu_\mu\bar{\nu}_\mu W^+_L W^-_L h $ with $c_{\Phi_1}=0$ }\\
\hline
$c_6$  &  $-2$  &  $-1$  &  $0$  &  $1$  &  $2$  \\
        \hline
1 TeV & $4.17 \times 10^{-9}$ & $1.57 \times 10^{-9}$ & $8.22 \times 10^{-10}$ & $1.93 \times 10^{-9}$ & $4.90 \times 10^{-9}$ \\
\hline
3 TeV & $1.79 \times 10^{-6}$ & $6.98 \times 10^{-7}$ & $3.71 \times 10^{-7}$ & $8.01 \times 10^{-7}$ & $2.00 \times 10^{-6}$ \\
\hline
5 TeV & $6.10 \times 10^{-6}$ & $2.43 \times 10^{-6}$ & $1.32 \times 10^{-6}$ & $2.74 \times 10^{-6}$ & $6.72 \times 10^{-6}$\\
\hline
10 TeV & $1.94 \times 10^{-5}$ & $7.98 \times 10^{-6}$ & $4.38 \times 10^{-6}$ & $8.74 \times 10^{-6}$ & $2.09 \times 10^{-5}$\\
\hline
14 TeV & $2.99 \times 10^{-5}$ & $1.25 \times 10^{-5}$ & $7.11 \times 10^{-6}$ & $1.36 \times 10^{-5}$ & $3.22 \times 10^{-5}$\\
\hline
30 TeV & $6.45 \times 10^{-5}$ & $2.82 \times 10^{-5}$ & $1.58 \times 10^{-5}$ & $2.95 \times 10^{-5}$ &$6.68 \times 10^{-5}$ \\
\hline
\end{tabular}
\caption{The cross section for $\mu^+\mu^-\to \nu_\mu\bar\nu_\mu W^+_L W^-_L h$ with $c_{\Phi_1}=0$ at different c.m. energies.
Five benchmark points of $c_6$ are displayed in different columns.
The cuts $m_{\nu\nu}>150$ GeV, $p_T(W,h)>150$ GeV are implemented to obtain these cross sections.}
\label{tab:cs_mm_wwh_c6}
\end{table}

\begin{table}[!tb]
\centering
\begin{tabular}{ |c|c|c|c|c|c|}
\hline
\multicolumn{6}{|c|}{Cross sections (pb) for $\mu^+ \mu^-\rightarrow \nu_\mu\bar{\nu}_\mu W^+_L W^-_L h $ with $c_6=0$ }\\
\hline
$c_{\Phi_1}$  &  $-2$  &  $-1$  &  $0$  & $1$  &  $2$  \\
\hline
1 TeV  & $6.23\times 10^{-7}$ & $6.70\times 10^{-7}$ & $7.69\times 10^{-7}$ & $9.27\times 10^{-7}$ & $1.14\times 10^{-6}$ \\
\hline
3 TeV  & $2.65\times 10^{-5}$ & $2.34\times 10^{-5}$ & $2.47\times 10^{-5}$ & $2.96\times 10^{-5}$ & $3.95\times 10^{-5}$ \\
\hline
5 TeV  & $8.85\times 10^{-5}$ & $6.71\times 10^{-5}$ & $6.43\times 10^{-5}$ & $8.13\times 10^{-5}$ & $1.17\times 10^{-4}$ \\
\hline
10 TeV & $4.09\times 10^{-4}$ & $2.22\times 10^{-4}$ & $1.71\times 10^{-4}$ & $2.61\times 10^{-4}$ & $4.72\times 10^{-4}$ \\
\hline
14 TeV & $8.63\times 10^{-4}$ & $3.95\times 10^{-4}$ & $2.56\times 10^{-4}$ & $4.38\times 10^{-4}$ & $9.39\times 10^{-4}$ \\
\hline
30 TeV & $4.97\times 10^{-3}$ & $1.60\times 10^{-3}$ &  $5.07\times 10^{-4}$ & $1.66\times 10^{-3}$ & $5.04\times 10^{-3}$ \\
\hline
\end{tabular}
\caption{
The same as~\autoref{tab:cs_mm_wwh_c6}, but for $c_{6}=0$ with five benchmark points of $c_{\Phi_1}$ at different c.m. energies. The cuts $m_{\nu\nu}>150$ GeV is implemented to obtain these cross sections
}
\label{tab:cs_mm_wwh_cphi1}
\label{tab:eevvw+0w-0h}
\end{table}

\begin{table}[!tb]
    \begin{center}
        \resizebox{\textwidth}{!}{
        \begin{tabular}{|c|c|c|c|c|}
        \hline\hline
            & \multicolumn{2}{c|}{$WWh$} & \multicolumn{2}{c|}{$hhh$}\\
        \hline
        Beams:  &  $\mu^+\mu^-$ & $pp$ & $\mu^+\mu^-$ & $pp$ \\
        \hline
        \multirow{4}{*}{Varying $c_6$:} &  $m_{\nu\nu}>150$ GeV & $m_{j_1j_2}>150$ GeV & $m_{\nu\nu}>150$ GeV & $m_{j_1j_2}>150$ GeV \\
                         &  $p_T(W,h)>150$ GeV & $p_T(W,h)>150$ GeV &  & $\eta_{j_1}\times\eta_{j_2}<0$ \\
                         &                     & $\eta_{j_1}\times\eta_{j_2}<0$ & & $|\Delta\eta_{j_1j_2}|>2.5$ \\
                         &                     & $|\Delta\eta_{j_1j_2}|>2.5$ & & \\
        \hline
        \multirow{3}{*}{Varying $c_{\Phi_1}$:} & $m_{\nu\nu}>150$ GeV & $m_{j_1j_2}>150$ GeV & $m_{\nu\nu}>150$ GeV & $m_{j_1j_2}>150$ GeV \\
                          &    & $\eta_{j_1}\times\eta_{j_2}<0$ & & $\eta_{j_1}\times\eta_{j_2}<0$ \\
                          &    & $|\Delta\eta_{j_1j_2}|>2.5$ & & $|\Delta\eta_{j_1j_2}|>2.5$ \\
        \hline\hline
        \end{tabular}}
    \end{center}
    \caption{The cuts used for processes with $WWh$ and $hhh$ final states at $pp$ and lepton colliders respectively.}
    \label{tab:cuts}
\end{table}

In this category, we considered following processes for hadron and lepton colliders:
\begin{align}
    p\,p&\to jj W^\pm W^\pm h,\\
    l^+\,l^-&\to \nu_l\bar{\nu}_l W^+W^- h,
\end{align}
where, as stated in last section, we choose same-sign W-pair for hadron collider to suppress the backgrounds.

The cross sections for $\mu^+\mu^-\to \nu_\mu \bar\nu_\mu W^+_L W^-_L h$ for different choices of $c_6$ and $c_{\Phi_1}$ are listed in~\autoref{tab:cs_mm_wwh_c6} and~\autoref{tab:cs_mm_wwh_cphi1} respectively. The cuts we imposed on the cross section calculation are listed in~\autoref{tab:cuts}. Note that we impose a slightly stronger cuts for the case in~\autoref{tab:cs_mm_wwh_c6} (as well as the case in~\autoref{tab:cs_pp_wwh_c6} below). As in this case, the enhancement due to $c_6$ is not large which is overwhelmed by the Sudakov logarithms from soft/collinear behavior in the SM cross section. We thus impose additional $p_T$ cuts on final states to avoid such soft/collinear regions.

From these tables, we find that the behavior of the cross section with respect to $c_6$ and $c_{\Phi_1}$ is similar to what we observed in~\autoref{sec:subprocess}. For $WWh$ final state, the dependence of the cross section on $c_6$ will be slightly weaker at higher energy, i.e. $\frac{\sigma(c_6=2)}{\sigma_{\rm SM}} = 5.96(4.23)$ for $\sqrt{s}=1(30)$ TeV. On the other hand, the cross section enhancement due to $c_{\Phi_1}$ will be stronger at higher energy, i.e. $\frac{\sigma(c_{\Phi_1}=2)}{\sigma_{\rm SM}} = 1.48(9.94)$ for $\sqrt{s}=1(30)$ TeV. Hence, by just measuring the total events, we will have stronger constraints on $c_{\Phi_1}$ than $c_6$ with a 14 TeV or even 30 TeV muon collider machine.

In the simulation we have chosen longitudinal polarizations for $W$ bosons in the final states. The reason is that the cross sections for summing over the polarizations of the final state $W^{\pm}$s are dominated by transverse polarizations, whereas deviation of Higgs self-couplings mainly modifies cross sections of all longitudinal vector bosons only. So in order to study the influence from high dimension operators, in practice, the longitudinal polarizations should be picked using some technics~\cite{Ballestrero:2017bxn,Ballestrero:2020qgv,De:2020iwq,Grossi:2020orx,Kim:2021gtv}. Detailed comparison between cross sections summing over polarizations of $W$ bosons in the final states and the ones with longitudinal $W$ bosons in the final states can be found in~\autoref{Sec:Appendix-2}.

\begin{table}[!tb]
    \centering
    \begin{tabular}{ |c|c|c|c|c|c|}
     \hline
     \multicolumn{6}{|c|}{Cross sections (pb) for $pp\rightarrow jj W^{\pm}_L W^{\pm}_L h $ with $c_{\Phi_1}=0$ }\\
     \hline
    $c_6$  &  $-2$  &  $-1$  &  $0$  &  $1$  &  $2$  \\
     \hline
    14 TeV  & $6.35\times 10^{-7}$ & $2.81\times 10^{-7}$ & $1.68\times 10^{-7}$ & $3.01\times 10^{-7}$ & $6.79\times 10^{-7}$ \\
     \hline
    27 TeV & $3.58\times 10^{-6}$ & $1.68\times 10^{-6}$ & $1.09\times 10^{-6}$ & $1.78\times 10^{-6}$ & $3.76\times 10^{-6}$ \\
     \hline
    100 TeV & $3.28\times 10^{-5}$& $1.82\times 10^{-5}$ & $1.38\times 10^{-5}$& $1.87\times 10^{-5}$ & $3.34\times 10^{-5}$ \\
     \hline
    \end{tabular}
    \caption{The cross section for $p\,p\to jj W^{\pm}_L W^{\pm}_L h$ with $c_{\Phi_1}=0$ at different c.m. energies.
    Five benchmark points of $c_6$ are displayed in different columns.
    The cuts $m_{jj}>150$ GeV, $p_T(W,h)>150$ GeV as well as the VBS selections ($ \eta_{j_1}\times\eta_{j_2} < 0 $ and $ |\Delta\eta_{j_1 j_2}| > 2.5 $) are implemented to obtain these cross sections.}
    \label{tab:cs_pp_wwh_c6}
\end{table}

\begin{table}[!tb]
    \centering
     \begin{tabular}{ |c|c|c|c|c|c|}
     \hline
     \multicolumn{6}{|c|}{Cross sections (pb) for $pp\rightarrow jj W^{\pm}_LW^{\pm}_L h $ with $c_6=0$ }\\
     \hline
    $c_{\Phi_1}$  &  $-2$  &  $-1$  &  $0$  &  $1$  &  $2$  \\
    \hline
    14 TeV  & $2.58\times 10^{-5}$ & $2.65\times 10^{-5}$ &  $2.88\times 10^{-5}$ & $3.29\times 10^{-5}$ & $3.90\times 10^{-5}$ \\
    \hline
    27 TeV  & $1.27\times 10^{-4}$ & $1.20\times 10^{-4}$ & $ 1.27\times 10^{-4}$ & $1.49\times 10^{-4}$ & $1.83\times 10^{-4}$ \\
    \hline
    100 TeV & $1.26\times 10^{-3}$ & $9.91\times 10^{-4}$ &  $9.70\times 10^{-4}$ & $1.23\times 10^{-3}$ & $1.72\times 10^{-3}$ \\
    \hline
    \end{tabular}
    \caption{
    The same as~\autoref{tab:cs_pp_wwh_c6}, but for $c_{6}=0$ with five benchmark points of $c_{\Phi_1}$ at different c.m. energies. The cuts $m_{jj}>150$ GeV and the VBS selections ($ \eta_{j_1}\times\eta_{j_2} < 0 $ and $ |\Delta\eta_{j_1 j_2}| > 2.5 $) are implemented to obtain these cross sections.
    }
    \label{tab:cs_pp_wwh_cphi1}
    \label{tab:ppjjw+0w-0h}
    \end{table}

The cross section for $p\,p\to jjW^{\pm}_LW^{\pm}_Lh$ with different choices of $c_6$ and $c_{\Phi_1}$ are listed in~\autoref{tab:cs_pp_wwh_c6} and~\autoref{tab:cs_pp_wwh_cphi1}. The relevant cuts applied on this process are also listed in~\autoref{tab:cuts}. The overall behavior of the cross section with respect to $c_6$ and $c_{\Phi_1}$ is similar to the case at muon collider, however, the sensitivity is weaker: $\frac{\sigma(c_6=2)}{\sigma_{\rm SM}} = 4.04(2.42)$ and $\frac{\sigma(c_{\Phi_1}=2)}{\sigma_{\rm SM}} \approx 1.35(1.77)$ for $\sqrt{s}=14(100)$ TeV.

\subsubsection*{Production of $hhh$}

In this category, the processes we considered at hadron and lepton colliders are:
\begin{align}
    p\,p&\to j j hhh,\\
    l^+l^-&\to \nu_l\bar\nu_l hhh.
\end{align}

\begin{table}[!tb]
    \centering
     \begin{tabular}{ |c|c|c|c|c|c|}
     \hline
     \multicolumn{6}{|c|}{Cross sections (pb) for $\mu^+ \mu^-\rightarrow \nu_\mu\bar{\nu}_\mu hhh $ with $c_{\Phi_1}=0$ }\\
     \hline
    $c_6$  &  $-2$  &  $-1$  &  $0$  &  $1$  &  $2$  \\
     \hline
     1 TeV  & $4.42 \times 10^{-8}$ & $1.06 \times 10^{-8}$ & $ 3.39\times 10^{-9}$ & $2.25 \times 10^{-8}$ & $6.76 \times 10^{-8}$ \\
     \hline
     3 TeV  & $1.93 \times 10^{-6}$ & $5.66 \times 10^{-7}$ & $ 2.78 \times 10^{-7}$ & $1.08 \times 10^{-6}$ & $2.94 \times 10^{-6}$ \\
     \hline
     5 TeV  & $4.91 \times 10^{-6}$ & $1.57 \times 10^{-6}$ & $ 9.50 \times 10^{-7}$ & $3.03 \times 10^{-6}$ & $7.74 \times 10^{-6}$ \\
      \hline
    10 TeV  & $1.25 \times 10^{-5}$ & $4.59 \times 10^{-6}$ & $ 3.46 \times 10^{-6}$ & $8.75 \times 10^{-6}$ & $1.95 \times 10^{-5}$ \\
     \hline
    14 TeV  & $1.80 \times 10^{-5}$ & $6.70 \times 10^{-6}$ &   $5.38\times 10^{-6}$ & $1.30 \times 10^{-5}$ & $2.88 \times 10^{-5}$ \\
     \hline
    30 TeV  & $3.50 \times 10^{-5}$ & $1.77 \times 10^{-5}$ & $1.41 \times 10^{-5}$ & $ 2.92 \times 10^{-5}$ & $5.42 \times 10^{-5}$ \\
     \hline
    \end{tabular}
    \caption{The cross section for $\mu^+\mu^-\to \nu_\mu\bar\nu_\mu hhh$ with $c_{\Phi_1}=0$ at different c.m. energies.
   Five benchmark points of $c_6$ are displayed in different columns.
    The cut $m_{\nu\nu}>150$ GeV is implemented to obtain these cross sections.}
    \label{tab:cs_mm_hhh_c6}
\end{table}

\begin{table}[!tb]
    \centering
    \begin{tabular}{ |c|c|c|c|c|c|}
     \hline
     \multicolumn{6}{|c|}{Cross sections (pb) for $\mu^+ \mu^-\rightarrow \nu_\mu\bar{\nu}_\mu hhh $ with $c_6=0$ }\\
     \hline
    $c_{\Phi_1}$  &  $-2$  &  $-1$ & $0$ &  $1$  &  $2$  \\
     \hline
     1 TeV  & $2.78 \times 10^{-8}$ & $1.08 \times 10^{-8}$ & $3.39\times10^{-9}$ & $5.56 \times 10^{-9}$ & $1.73 \times 10^{-8}$ \\
     \hline
     3 TeV  & $3.01 \times 10^{-6}$ & $1.11 \times 10^{-6}$ & $2.78\times10^{-7}$ & $5.43 \times 10^{-7} $ & $1.89 \times 10^{-6}$ \\
     \hline
     5 TeV  & $1.33 \times 10^{-5}$ & $4.47 \times 10^{-6}$ & $9.50\times10^{-7}$ & $2.38 \times 10^{-6} $ & $8.76 \times 10^{-6}$ \\
     \hline
    10 TeV  & $7.83 \times 10^{-5}$ & $2.38 \times 10^{-5}$ & $3.46\times10^{-6}$ & $1.50 \times 10^{-5}$ & $5.97 \times 10^{-5}$ \\
     \hline
    14 TeV  & $1.77 \times 10^{-4}$ & $4.97 \times 10^{-5}$ & $5.38\times10^{-6}$ & $ 3.73 \times 10^{-5}$ & $1.44 \times 10^{-4}$ \\
     \hline
    30 TeV  & $1.07 \times 10^{-3}$ & $2.77 \times 10^{-4}$ & $1.41\times10^{-5}$ & $ 2.44 \times 10^{-4}$ & $9.86 \times 10^{-4}$ \\
     \hline
    \end{tabular}
     \caption{
     The same as~\autoref{tab:cs_mm_hhh_c6}, but for $c_{6}=0$ with five benchmark points of $c_{\Phi_1}$ at different c.m. energies.
    }
    \label{tab:cs_mm_hhh_cphi1}
     \label{tab:eevvhhh}
\end{table}
The cross section for $\mu^+\mu^-\to\nu_\mu\bar\nu_\mu hhh$ for different choices of $c_6$ and $c_{\Phi_1}$ are listed in~\autoref{tab:cs_mm_hhh_c6} and~\autoref{tab:cs_mm_hhh_cphi1}. The cuts we imposed on the process are also listed in~\autoref{tab:cuts}. The cross sections for $hhh$ production are slightly smaller than that of $WWh$ production, while the sensitivity on $c_{\Phi_1}$ from $hhh$ channel is much stronger than that in $WWh$ channel as we can see from~\autoref{tab:cs_mm_hhh_cphi1}, $\frac{\sigma(c_{\Phi_1}=2)}{\sigma_{\rm SM}} \approx 70$ at $\sqrt{s}=30$ TeV. However, the enhancement due to $c_6$ is moderate: $\frac{\sigma(c_6=2)}{\sigma_{\rm SM}} \approx 4$ at $\sqrt{s}=30$ TeV. The cross section and its dependence on $c_6$ and $c_{\Phi_1}$ for $p\,p\to jjhhh$ are listed in~\autoref{tab:cs_pp_hhh_c6} and~\autoref{tab:cs_pp_hhh_cphi1}. Similar to muon collider case, the sensitivities on both $c_6$ and $c_{\Phi_1}$ are stronger from $hhh$ production than that from $WWh$ production. At $\sqrt{s}=100$ TeV, we have $\frac{\sigma(c_6=2)}{\sigma_{\rm SM}}\approx8$ and $\frac{\sigma(c_{\Phi_1})}{\sigma_{\rm SM}}\approx18$.

\begin{table}[!tb]
\centering
\begin{tabular}{ |c|c|c|c|c|c|}
\hline
\multicolumn{6}{|c|}{Cross sections (pb)  for $pp\rightarrow jj hhh $ with $c_{\Phi_1}=0$}\\
\hline
$c_6$  &  $-2$  &  $-1$  &  $0$  &  $1$  &  $2$  \\
\hline
14 TeV  & $1.99\times 10^{-6}$ & $5.77\times 10^{-7}$ & $ 2.97\times 10^{-7}$ & $1.16\times 10^{-6}$ & $3.12\times 10^{-6}$ \\
    \hline
27 TeV  & $9.46\times 10^{-6}$ & $2.93\times 10^{-6}$ & $ 1.50\times 10^{-6}$ & $5.48\times 10^{-6}$ & $1.45\times 10^{-5}$ \\
    \hline
100 TeV & $7.91\times 10^{-5}$ & $2.65\times 10^{-5}$ & $ 1.48\times 10^{-5} $ & $4.30\times 10^{-5}$ & $ 1.13\times 10^{-4} $ \\
    \hline
\end{tabular}
\caption{The cross section for $p\,p\to jj hhh$ with $c_{\Phi_1}=0$ at different c.m. energies.
Five benchmark points of $c_6$ are displayed in different columns.
The cut $m_{j_1j_2}>150$ GeV and VBS selections ($ \eta_{j_1}\times\eta_{j_2} < 0 $ and $ |\Delta\eta_{j_1 j_2}| > 2.5 $) are implemented to obtain these cross sections.}
\label{tab:cs_pp_hhh_c6}
\end{table}

\begin{table}[!tb]
\centering
\begin{tabular}{ |c|c|c|c|c|c|}
\hline
\multicolumn{6}{|c|}{Cross sections (pb) for $pp\rightarrow jj hhh $ with $c_6=0$}\\
\hline
$c_{\Phi_1}$  &  $-2$  &  $-1$ & $0$ &  $1$  &  $2$  \\
\hline
14 TeV & $2.94\times 10^{-6}$ & $1.10\times 10^{-6}$ & $2.97\times10^{-7}$ & $ 5.33\times 10^{-7}$ & $1.83\times 10^{-6}$ \\
\hline
27 TeV & $1.97\times 10^{-5}$ & $6.99\times 10^{-6}$ & $1.50\times10^{-6}$ & $ 3.75\times 10^{-6}$ & $1.35\times 10^{-5}$ \\
\hline
100 TeV& $3.31\times 10^{-4}$ & $1.02\times 10^{-4}$ & $1.48\times10^{-5}$ & $ 7.04\times 10^{-5}$ & $2.69\times 10^{-4}$ \\
\hline
\end{tabular}
\caption{
The same as~\autoref{tab:cs_pp_hhh_c6}, but for $c_{6}=0$ with five benchmark points of $c_{\Phi_1}$ at different c.m. energies.
    }
\label{tab:cs_pp_hhh_cphi1}
\label{tab:ppjjhhh}
\end{table}

\subsection{Results and Prospects}
\label{sec:simulation_plot}

\begin{figure}[!tb]
\centering
\includegraphics[width=0.48 \textwidth]{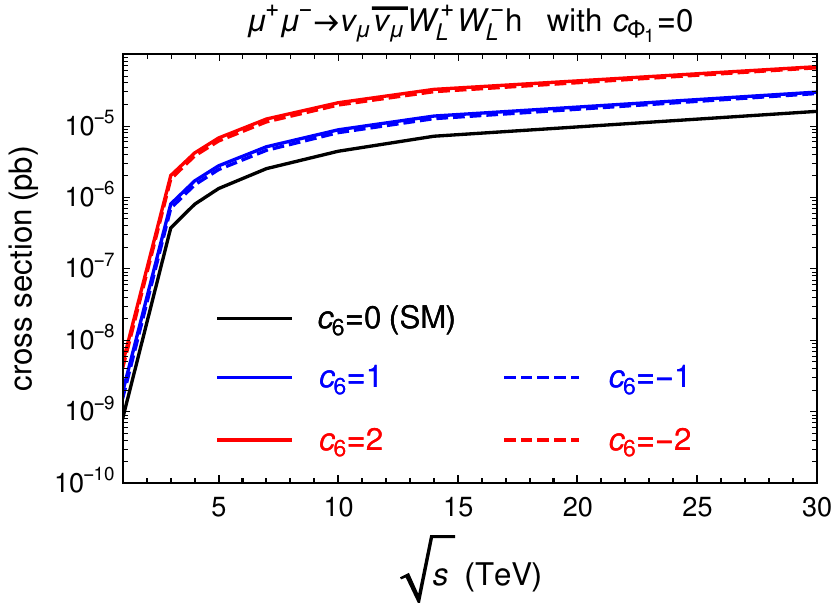}
\includegraphics[width=0.48 \textwidth]{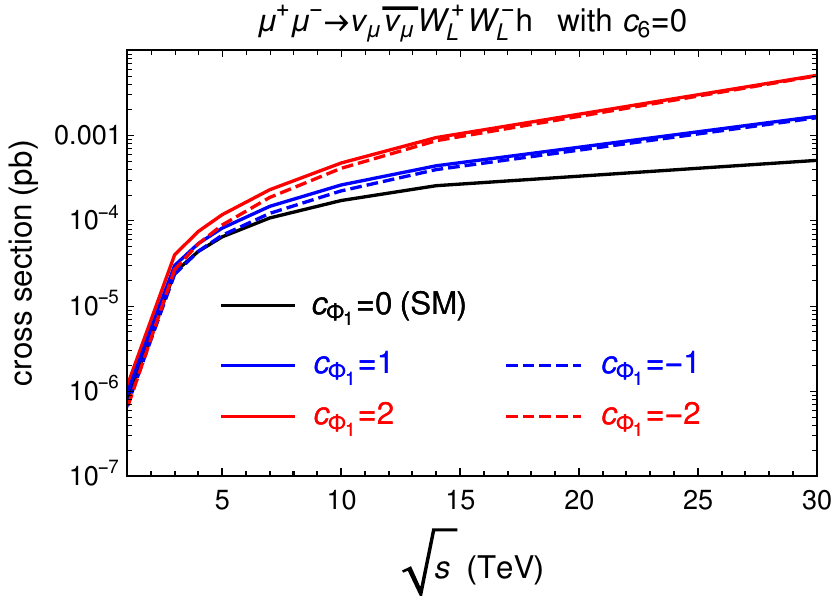}\\
\includegraphics[width=0.48 \textwidth]{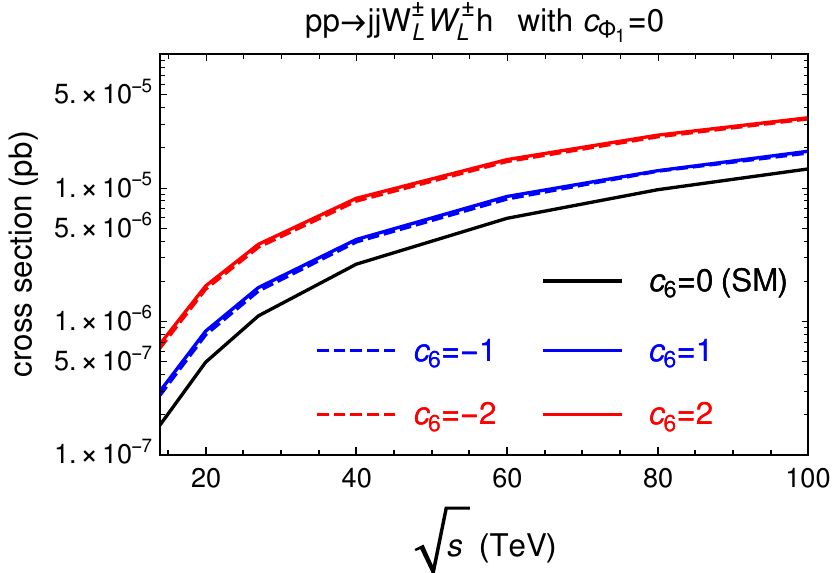}
\includegraphics[width=0.48 \textwidth]{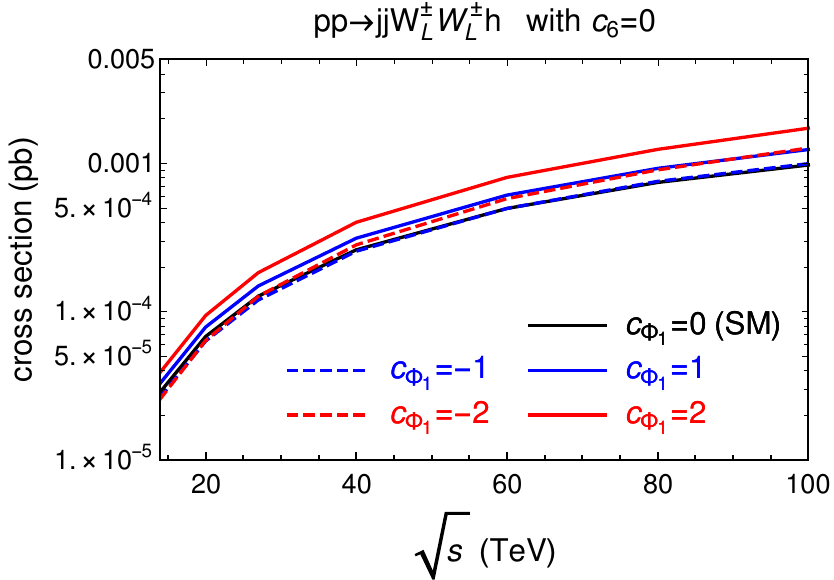}
\caption{The cross section for $WWh$ production at muon collider (upper panels) and hadron collider (lower panels) as function of $\sqrt{s}$. }
\label{fig:cs_wwh}
\label{fig:wwh_had}
\end{figure}

\begin{figure}[!tb]
\centering
\includegraphics[width=0.48 \textwidth]{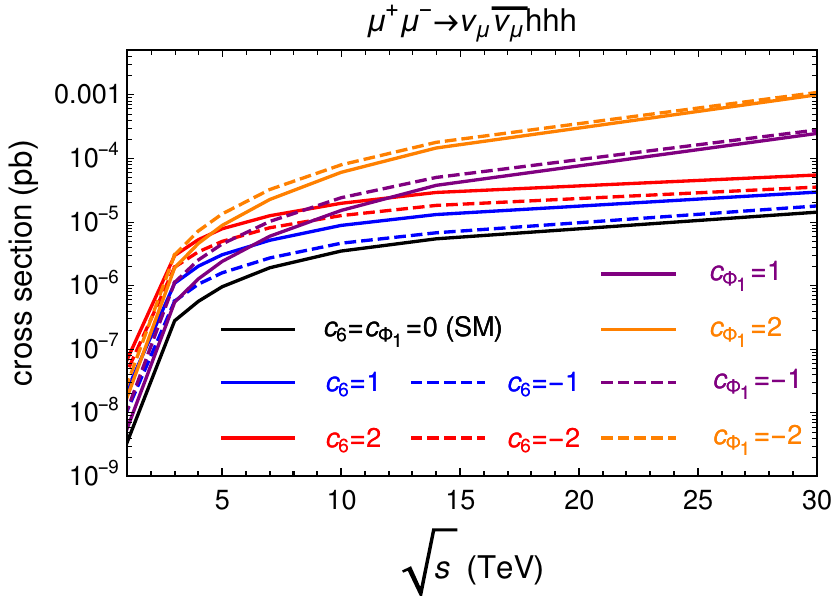}
\includegraphics[width=0.48 \textwidth]{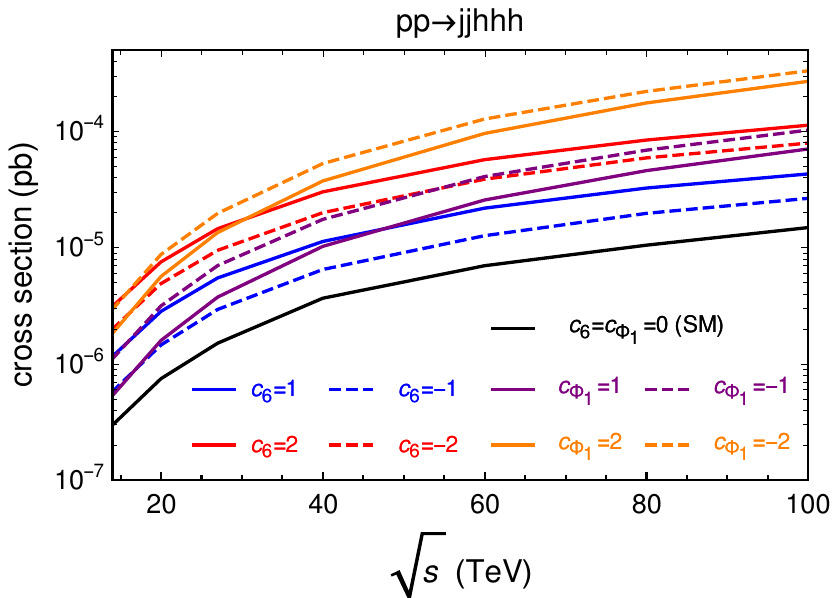}
\caption{The vary of cross sections for $c_6 =\pm 1, \pm 2$ with $c_{\Phi_1}=0$ and $c_{\Phi_1} =\pm 1, \pm 2$ with $c_6 =0$ for $\mu^+\mu^-\rightarrow\nu_{\mu}\overline{\nu}_{\mu}hhh$ from $\sqrt{s}= 1$ to $30$ TeV (left panel) and $pp\rightarrow jjhhh$ from $\sqrt{s}= 14$ to $100$ TeV (right panel).}
\label{fig:hhh}
\end{figure}

Based on the cross section results in~\autoref{sec:simulation_data}, in~\autoref{fig:cs_wwh} and~\autoref{fig:hhh}, we show the cross section as function of the c.m. energies for both $WWh$ and $hhh$ productions where the black curve is for the SM case and different colors (solid or dashed) are for the cases either $c_6$ or $c_{\Phi_1}$ is non-zero. Notice that the cross section increases logarithmically as the main production mechanism is VBS. All these processes provide some sensitivities on both $c_6$ and $c_{\Phi_1}$ as can be seen by comparing the colored curves with the SM one.

\begin{figure}[!tb]
\centering
\includegraphics[width=0.48\textwidth]{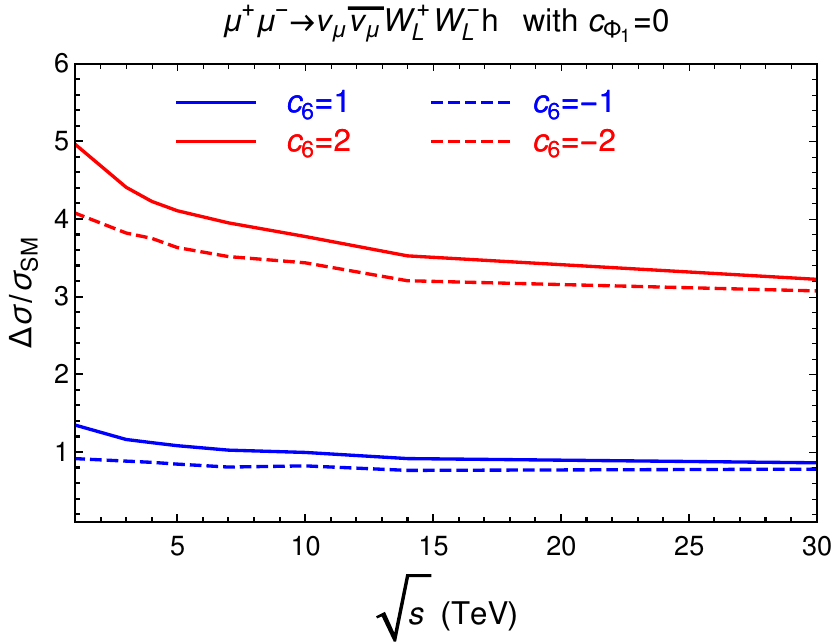}
\includegraphics[width=0.48\textwidth]{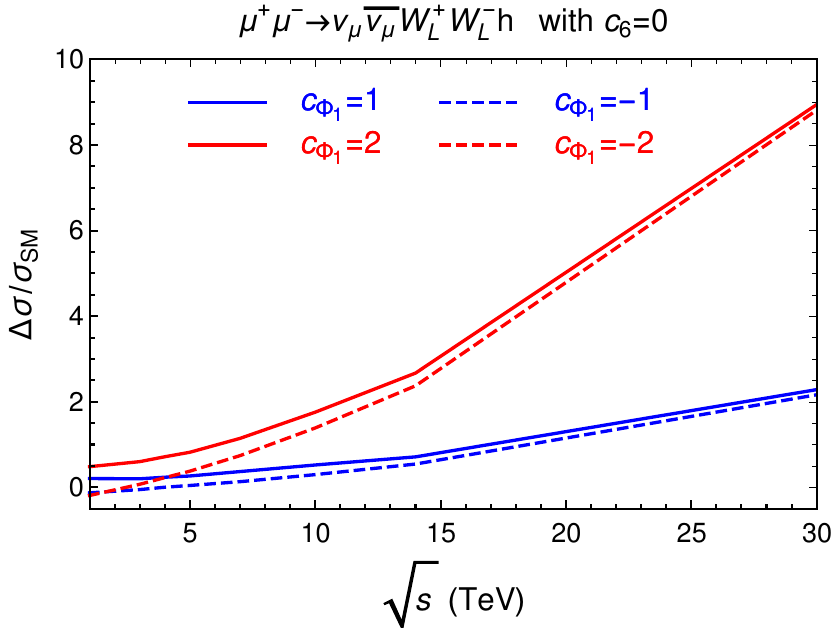}\\
\includegraphics[width=0.48\textwidth]{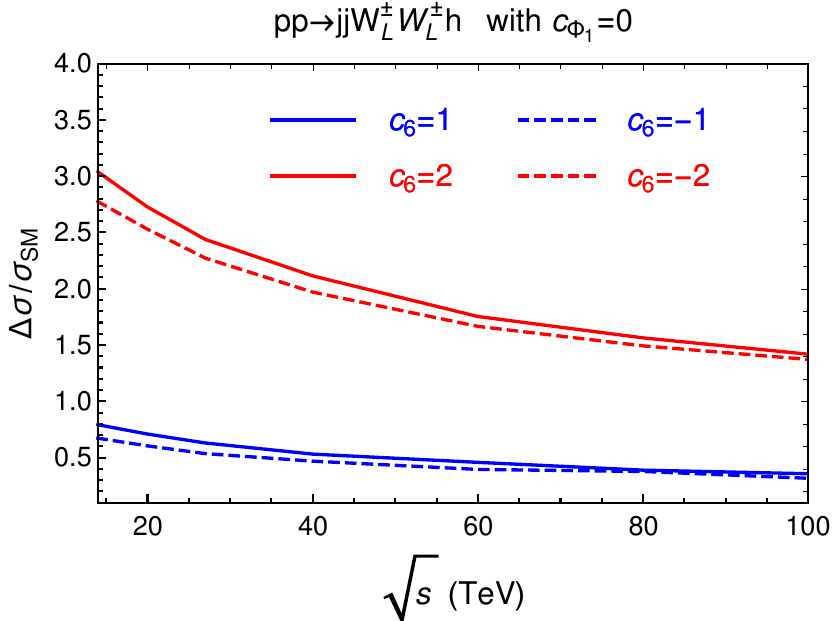}
\includegraphics[width=0.48\textwidth]{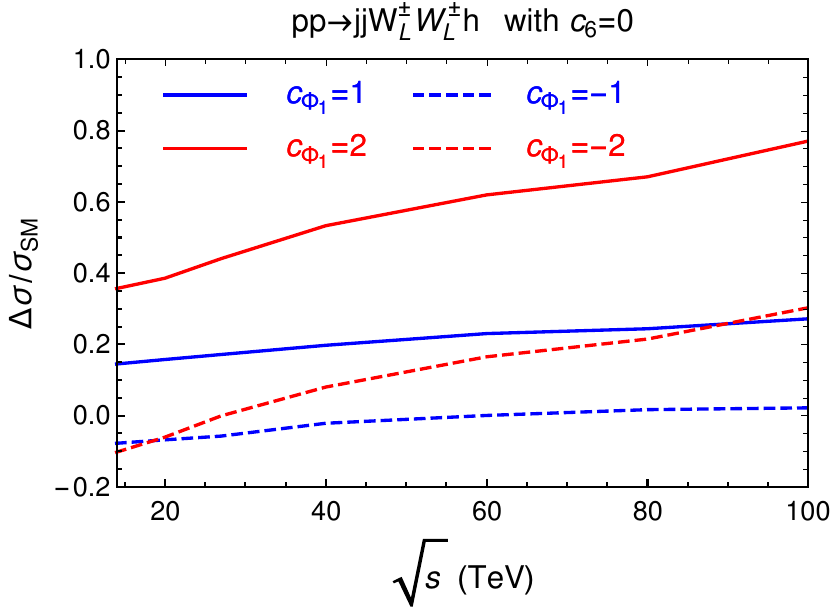}
\caption{The $\Delta\sigma /\sigma_{SM}$ for $c_6 =\pm 1, \pm 2$ with $c_{\Phi_1}=0$ and $c_{\Phi_1} =\pm 1, \pm 2$ with $c_6 =0$ for $\mu^+\mu^-\rightarrow\nu_{\mu}\overline{\nu}_{\mu}WWh$ from $\sqrt{s}= 1$ to $30$ TeV (upper panel) and $pp\rightarrow jjWWh$ from $\sqrt{s}= 14$ to $100$ TeV (lower panel) where $\Delta\sigma\equiv\sigma_{tot}-\sigma_{SM}$. }
\label{fig:ratio_wwh}
\end{figure}

\begin{figure}[h]
\centering
\includegraphics[width=0.48 \textwidth]{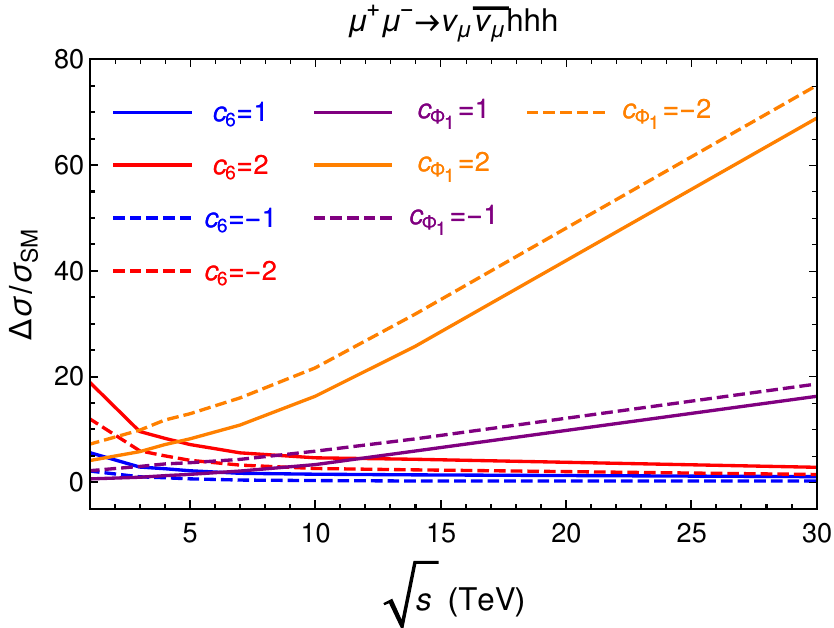}
\includegraphics[width=0.48 \textwidth]{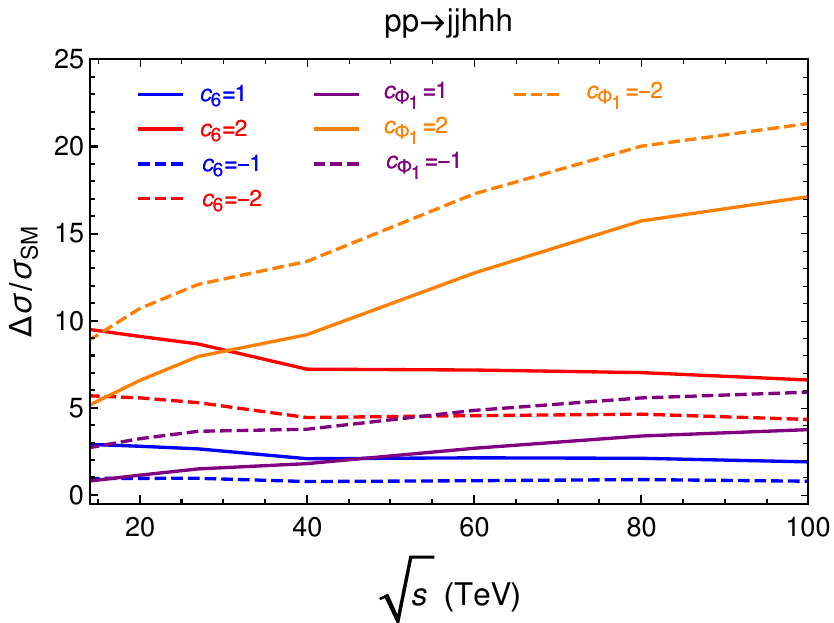}
\caption{
The $\Delta\sigma /\sigma_{SM}$ for $c_6 =\pm 1, \pm 2$ with $c_{\Phi_1}=0$ and $c_{\Phi_1} =\pm 1, \pm 2$ with $c_6 =0$ for $\mu^+\mu^-\rightarrow\nu_{\mu}\overline{\nu}_{\mu}hhh$ from $\sqrt{s}= 1$ to $30$ TeV (left panel) and $pp\rightarrow jjhhh$ from $\sqrt{s}= 14$ to $100$ TeV (right panel) where $\Delta\sigma\equiv\sigma_{tot}-\sigma_{SM}$. }
\label{fig:ratio_hhh}
\end{figure}

In~\autoref{fig:ratio_wwh} and~\autoref{fig:ratio_hhh}, we show the cross section difference induced by $\mathcal{O}_6$ and $\mathcal{O}_{\Phi_1}$ compared with the SM one for $WWh$ and $hhh$ production respectively. In general, the sensitivity at lepton collider will be stronger than that at hadron collider. Further, the sensitivity on $c_6$ ($\mathcal{O}_6$) decreases as energy increases, as we have indicated in previous sections, due to the logarithmical enhancement of the SM cross section. On the other hand, the sensitivity on $c_{\Phi_1}$ ($\mathcal{O}_{\Phi_1}$) increases at high energy due to the momentum dependence in the $\mathcal{O}_{\Phi_1}$ operator.

\begin{table}[!tb]
    \centering
     \begin{tabular}{ |c|c|c|c|c|c|}
     \hline
     \multicolumn{6}{|c|}{ Allowed region from the $\mu^+\mu^-\rightarrow\nu_{\mu}\overline{\nu}_{\mu}W^+_L W^-_L h$ process }\\
     \hline
     \multirow{2}{*}{$\sqrt{s}$ (TeV)} & \multirow{2}{*}{$\mathcal{L}$ ($\rm ab^{-1}$)} & \multicolumn{2}{c|}{$c_6$ ($c_{\Phi_1}=0$) x-sec only} & \multicolumn{2}{c|}{$c_{\Phi_1}$ ($c_6=0$) x-sec only} \\
     \cline{3-6}
     &  & $1\sigma$ & $2\sigma$ & $1\sigma$ & $2\sigma$ \\
     \hline
     1 & 1.2 & $-$ & $-$ & $-$ & $-$\\
     \hline
     3 & 4.4 & $-$ & $-$ & $[-2.54,0.71]$ & $[-3.07,1.18]$\\
     \hline
     5 & 12 & $\left[ -0.65, 0.53\right]$ & $\left[ -0.89, 0.77\right]$ & $[-1.02,0.30]$ & $[-1.22,0.50]$ \\
     \hline
    10 & 20 & $\left[ -0.44, 0.35\right]$ & $\left[ -0.60, 0.51\right]$ & $[-0.43,0.13]$ & $[-0.50,0.22]$\\
     \hline
    14 & 33 & $\left[ -0.37, 0.28\right]$ & $\left[ -0.50, 0.41\right]$ & $[-0.23,0.10]$ & $[-0.29,0.15]$ \\
     \hline
    30 & 100 & $\left[ -0.23, 0.18\right]$ & $\left[ -0.31, 0.26\right]$ & $[-0.07,0.04]$ & $[-0.09,0.06]$ \\
     \hline
    \end{tabular}
    \caption{ The range of $c_6$ with $c_{\Phi_1}=0$ and $c_{\Phi_1}$ with $c_6 = 0$ for $\mu^+\mu^-\rightarrow\nu_{\mu}\overline{\nu}_{\mu}W^+_L W^-_L h$ process at lepton colliders with various benchmarks $\sqrt{s}$ and $\mathcal{L}$. The $1\sigma$ and $2\sigma$ allowed regions on $c_6$ and $c_{\Phi_1}$ are calculated from the definition in~\autoref{eq:sig} which only rely on the size of cross sections. The notation ``$-$'' means either $N_{\rm SM} < 1$ or $|N_{\rm BSM}-N_{\rm SM}| < 1$.}
    \label{tab:test1}
\end{table}

\begin{table}[!tb]
    \centering
     \begin{tabular}{ |c|c|c|c|c|c|}
     \hline
     \multicolumn{6}{|c|}{ Allowed region from the $\mu^+\mu^-\rightarrow\nu_{\mu}\overline{\nu}_{\mu}hhh$ process }\\
     \hline
     \multirow{2}{*}{$\sqrt{s}$ (TeV)} & \multirow{2}{*}{$\mathcal{L}$ ($\rm ab^{-1}$)} & \multicolumn{2}{c|}{$c_6$ ($c_{\Phi_1}=0$) x-sec only} & \multicolumn{2}{c|}{$c_{\Phi_1}$ ($c_6=0$) x-sec only} \\
     \cline{3-6}
     &  & $1\sigma$ & $2\sigma$ & $1\sigma$ & $2\sigma$ \\
     \hline
     1 & 1.2 & $-$ & $-$ & $-$ & $-$ \\
     \hline
     3 & 4.4 & $-$ & $-$ & $-$ & $-$ \\
     \hline
     5 & 12 & $\left[ -1.00, 0.46\right]$ & $\left[ -1.27, 0.73\right]$ & $[-0.34,0.76]$ & $[-0.54,0.96]$ \\
     \hline
    10 & 20 & $\left[ -0.95, 0.30\right]$ & $\left[ -1.14, 0.50\right]$ & $[-0.14,0.41]$ & $[-0.23,0.50]$ \\
     \hline
    14 & 33 & $\left[ -0.93, 0.21\right]$ & $\left[ -1.08, 0.37\right]$ & $[-0.10,0.24]$ & $[-0.16,0.30]$ \\
     \hline
    30 & 100 & $\left[ -0.75, 0.12\right]$ & $\left[ -0.84, 0.21\right]$ & $[-0.04,0.10]$ & $[-0.06,0.12]$ \\
     \hline
    \end{tabular}
    \caption{ The same as~\autoref{tab:test1} but for $\mu^+ \mu^-\rightarrow \nu_\mu\bar{\nu}_\mu hhh $ process at lepton colliders. }
  \label{tab:test2}
\end{table}

Based on the deviation in the cross section due to $\mathcal{O}_6$ and $\mathcal{O}_{\Phi_1}$ operators contributions, we can estimate the allowed region of $c_6$ and $c_{\Phi_1}$ under some simplified assumptions.
First, the full signal to background analysis is beyond our scope~\footnote{However, we'd like to comment that the background at lepton collider can be well controlled, as the cut on invariant mass of all final states can help to remove most of the background due to the VBS nature of our signal process. While at hadron collider, we have chosen same-sign $W$ boson production, as the specific 2 same-sign lepton signals will also suppress the possible background at hadron collider.}. In our analysis, we simply follow the strategy used in Ref.~\cite{Chiesa:2020awd} to count the deviation of signal events from SM ones. Further, we define the significance of non-SM events over SM ones as:
\begin{equation}
\text{significance}\sim\frac{|N_{\rm BSM}-N_{\rm SM}|}{\sqrt{N_{\rm SM}}}.
\label{eq:sig}
\end{equation}
where $N_{\rm SM}$ and $N_{\rm BSM}$ are the event number of SM processes and the case with $\mathcal{O}_6$ or $\mathcal{O}_{\Phi_1}$ operator respectively.
For $\mu^+\mu^-\rightarrow\nu_{\mu}\overline{\nu}_{\mu}W^+_L W^-_L h$ process, $N_i\equiv\sigma_i(\mu^+\mu^-\rightarrow\nu_{\mu}\overline{\nu}_{\mu}W^+_L W^-_L h)\times\left[{\rm BR}(h\rightarrow b\overline{b})\right]\times\left[ {\rm BR}(W^{\pm}\rightarrow \text{all})\right]^2\times \mathcal{L} $ where $\mathcal{L}$ is the integrated luminosity.
Here we assume that all of leptonic, semileptonic and hadronic decay modes from opposite sign $W$ boson pair can be detected at future muon collider experiments.
For $pp\rightarrow jjW^{\pm}_L W^{\pm}_L h$ process, $N_i\equiv\sigma_i(pp\rightarrow jjW^{\pm}_L W^{\pm}_L h)\times\left[{\rm BR}(h\rightarrow b\overline{b})\right]\times\left[{\rm BR}(W^{\pm}\rightarrow l^{\pm}\nu)\right]^2\times \mathcal{L} $, where we only focus on the leptonic decay modes of same sign $W$ boson pair at hadron colliders to avoid possible huge SM backgrounds. For the $hhh$ processes, $N_i\equiv\sigma_i\times\left[{\rm BR}(h\rightarrow b\overline{b})\right]^3\times\mathcal{L} $ where $\sigma_i$ is the production cross section for $\mu^+ \mu^-\rightarrow \nu_\mu\bar{\nu}_\mu hhh$ or $pp\rightarrow jjhhh$ process at muon colliders or pp colliders.
We also require both $N_{\rm SM} > 1$ and $|N_{\rm BSM}-N_{\rm SM}| > 1$ in our analysis.
The allowed region for $c_6$ and $c_{\Phi_1}$ at 1- and 2-$\sigma$ level are summarized in~\autoref{tab:test1}-\autoref{tab:test4}. Note that several entries of the allowed regions for $c_{\Phi_1}$ are beyond $[-2,2]$, which should be treated with caution. As in our analysis, we only calculate the cross section within $[-2,2]$ for both $c_6$ and $c_{\Phi_1}$, the interpolation is only valid within this region.

\begin{table}[!tb]
    \centering
     \begin{tabular}{ |c|c|c|c|c|c|}
     \hline
     \multicolumn{6}{|c|}{ Allowed region from the $pp\rightarrow jjW^{\pm}_L W^{\pm}_L h$ process }\\
     \hline
     \multirow{2}{*}{$\sqrt{s}$ (TeV)} & \multirow{2}{*}{$\mathcal{L}$ ($\rm ab^{-1}$)} & \multicolumn{2}{c|}{$c_6$ ($c_{\Phi_1}=0$) x-sec only} & \multicolumn{2}{c|}{$c_{\Phi_1}$ ($c_6=0$) x-sec only} \\
     \cline{3-6}
     &  & $1\sigma$ & $2\sigma$ & $1\sigma$ & $2\sigma$ \\
     \hline
     14 & 3   & $-$ & $-$ & $[-8.04,3.02]$ & $[-11.34,4.62]$ \\
     \hline
     27 & 15  & $-$ & $-$ & $[-2.76,0.87]$ & $[-3.29,1.46]$ \\
     \hline
    100 & 30 & $\left[ -0.99, 0.92\right]$ & $\left[ -1.38, 1.32\right]$ & $[-1.08,0.23]$ & $[-1.24,0.40]$ \\
     \hline
    \end{tabular}
    \caption{ The same as~\autoref{tab:test1} but for $pp\rightarrow jjW^{\pm}_L W^{\pm}_L h$ process at hadron colliders. }
    \label{tab:test3}
\end{table}

\begin{table}[!tb]
    \centering
     \begin{tabular}{ |c|c|c|c|c|c|}
     \hline
     \multicolumn{6}{|c|}{ Allowed region from the $pp\rightarrow jjhhh$ process }\\
     \hline
     \multirow{2}{*}{$\sqrt{s}$ (TeV)} & \multirow{2}{*}{$\mathcal{L}$ ($\rm ab^{-1}$)} & \multicolumn{2}{c|}{$c_6$ ($c_{\Phi_1}=0$) x-sec only} & \multicolumn{2}{c|}{$c_{\Phi_1}$ ($c_6=0$) x-sec only} \\
     \cline{3-6}
     &  & $1\sigma$ & $2\sigma$ & $1\sigma$ & $2\sigma$ \\
     \hline
     14 & 3   & $-$ & $-$ & $-$ & $-$ \\
     \hline
     27 & 15  & $\left[ -0.80, 0.33\right]$ & $\left[ -1.00, 0.53\right]$ & $[-0.19,0.79]$ & $[-0.33,0.90]$ \\
     \hline
    100 & 30 & $\left[ -0.55, 0.14\right]$ & $\left[ -0.65, 0.24\right]$ & $[-0.07,0.30]$ & $[-0.13,0.35]$ \\
     \hline
    \end{tabular}
    \caption{ The same as~\autoref{tab:test1} but for $pp\rightarrow jjhhh$ process at hadron colliders. }
  \label{tab:test4}
\end{table}

\begin{figure}
    \centering
    \includegraphics[width=1.0\textwidth]{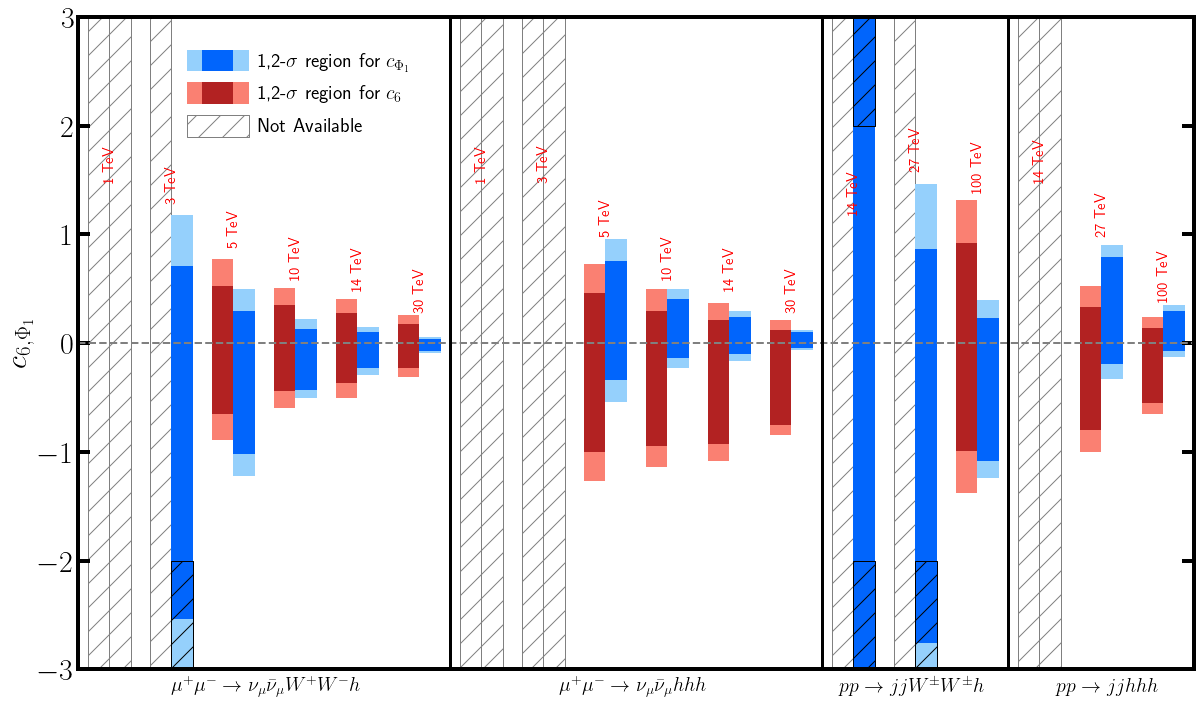}
    \caption{The allowed region for $c_6$ (red) and $c_{\Phi_1}$ (blue) from different channels. The darker color indicates the 1-$\sigma$ region, while lighter one indicates the 2-$\sigma$ region. The hatched region are not available either due to low event rate or beyond $[-2,2]$.}
    \label{fig:allowed_region_c6_cphi}
\end{figure}

The allowed regions for $c_6$ (red) and $c_{\Phi_1}$ (blue) are also shown in~\autoref{fig:allowed_region_c6_cphi} where darker color indicates the 1-$\sigma$ region, while lighter one indicates the 2-$\sigma$ region. We also denote those channels that cannot provide enough event rate as hatched region. It is clear from this plot that, in general, high energy muon collider is more powerful than HE-LHC in constraining $c_6$ and $c_{\Phi_1}$. Furthermore, both $WWh$ and $hhh$ production are more sensitive to $c_{\Phi_1}$ than $c_6$.  However, for both $c_6$ and $c_{\Phi_1}$, we will have higher sensitivities with higher energy at both lepton and hadron colliders thanks to the increase of  cross sections.

\begin{figure}
    \centering
    \includegraphics[width=0.9\textwidth]{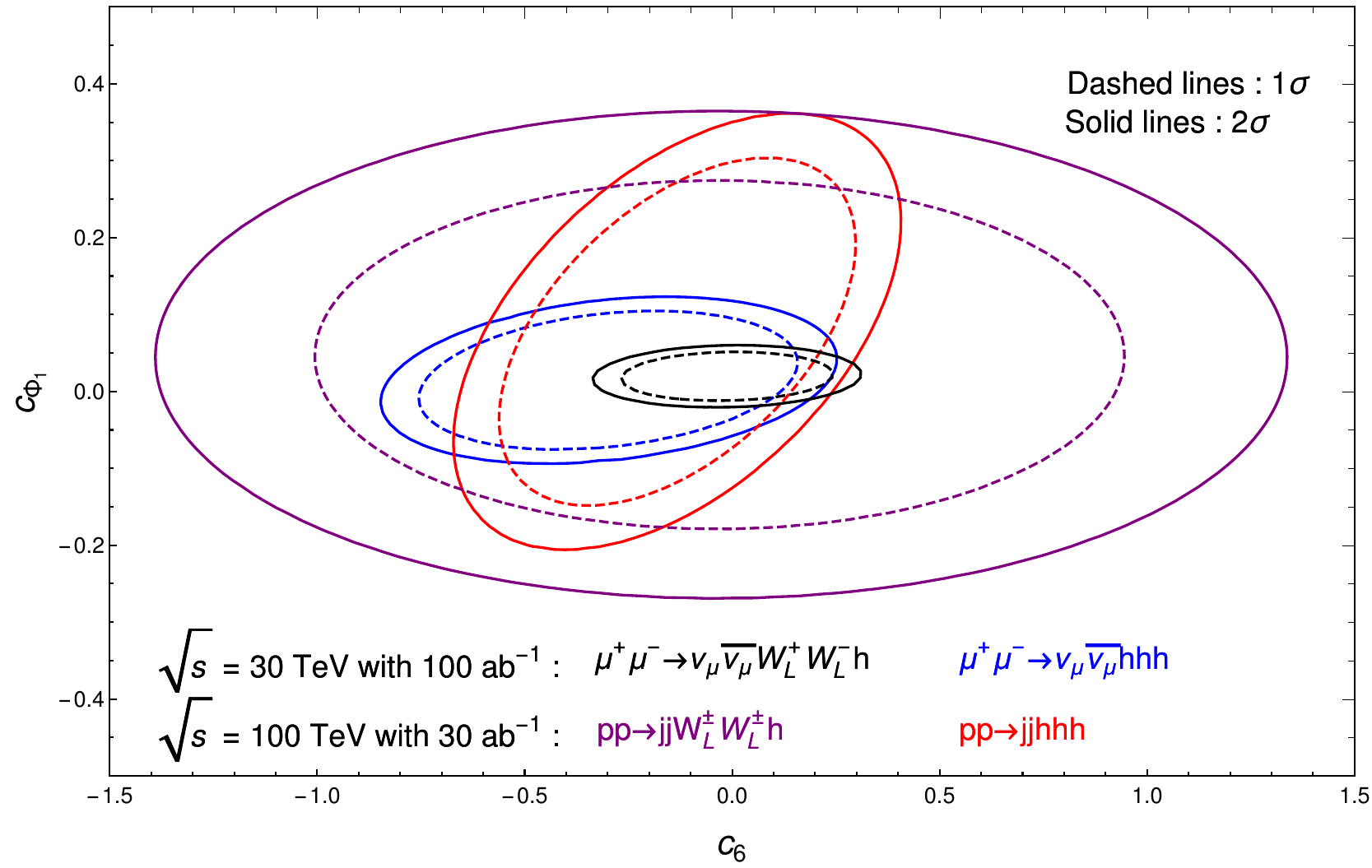}
    \caption{The allowed parameter space on the $ (c_6, c_{\Phi_1}) $ plane for 30 TeV muon collider with ${\mathcal L} = 100\ {\rm ab}^{-1}$ and 100 TeV hadron collider with ${\mathcal L} = 30\ {\rm ab}^{-1}$. The dashed (solid) lines represent 1-$\sigma$ (2-$\sigma$) allowed regions and four channels are labelled with $ \mu^+\mu^-\rightarrow\nu_{\mu}\overline{\nu_{\mu}}W^+_LW^-_Lh $ (black), $ \mu^+\mu^-\rightarrow\nu_{\mu}\overline{\nu_{\mu}}hhh $ (blue), $ p p\rightarrow j j W^{\pm}_LW^{\pm}_Lh $ (purple), and $ p p\rightarrow j j hhh $ (red).}
    \label{fig:2D_plot}
\end{figure}

We further study the allowed parameter space on the $ (c_6, c_{\Phi_1}) $ plane for 30 TeV muon collider with ${\mathcal L} = 100\ {\rm ab}^{-1}$ and 100 TeV hadron collider with ${\mathcal L} = 30 \ {\rm ab}^{-1}$ in~\autoref{fig:2D_plot}. First, the dashed (solid) lines represent 1-$\sigma$ (2-$\sigma$) allowed regions. Then, we include four channels with $ \mu^+\mu^-\rightarrow\nu_{\mu}\overline{\nu_{\mu}}W^+_LW^-_Lh $ (black), $ \mu^+\mu^-\rightarrow\nu_{\mu}\overline{\nu_{\mu}}hhh $ (blue), $ p p\rightarrow j j W^{\pm}_LW^{\pm}_Lh $ (purple), and $ p p\rightarrow j j hhh $ (red). For comparison purpose, we uniformly apply the stronger cuts in~\autoref{tab:cuts} for both $ \mu^+\mu^-\rightarrow\nu_{\mu}\overline{\nu_{\mu}}W^+_LW^-_Lh $ and $ p p\rightarrow j j W^{\pm}_LW^{\pm}_Lh $. We can find the most stringent constraint on the $ (c_6, c_{\Phi_1}) $ plane comes from
the process $ \mu^+\mu^-\rightarrow\nu_{\mu}\overline{\nu_{\mu}}W^+_LW^-_Lh $. However, $ \mu^+\mu^-\rightarrow\nu_{\mu}\overline{\nu_{\mu}}hhh $ and $ p p\rightarrow j j hhh $ processes can still help us to cover some parameter space on the $ (c_6, c_{\Phi_1}) $ plane which the process $ \mu^+\mu^-\rightarrow\nu_{\mu}\overline{\nu_{\mu}}W^+_LW^-_Lh $ cannot reach.

$2\rightarrow 3$ VBS processes are not the only channels to study the measurement of Higgs self-couplings at colliders. At the LHC, and future hadron colliders, di-Higgs production through gluon fusion remains the dominant channel~\cite{Sirunyan:2020xok,ATLAS:2021jki,Lu:2015jza,Adhikary:2017jtu,Goncalves:2018qas,Chang:2018uwu,Cepeda:2019klc,Cheung:2020xij}. Whereas at future lepton colliders, di-Higgs production through $2\rightarrow 2$ VBS, i.e. $VV\rightarrow hh$, become dominant~\cite{Robson:2018zje,Roloff:2019crr,Buttazzo:2018qqp, Costantini:2020stv,Han:2020pif}.   In terms of Wilson coefficients,  both $c_6$ and $c_{\Phi_1}$  can  also be measured with either $gg\rightarrow hh$ or $VV\rightarrow hh$. Furthermore,  $c_{\Phi_1}$ can be measured with $2\rightarrow 1$ vector boson fusion, $2\rightarrow 2$ VBS with gauge boson final states and etc.  See also~\cite{DiMicco:2019ngk} and~\cite{deBlas:2019rxi} for overall reviews.

A comparison between muon colliders and other future colliders was done in~\cite{Costantini:2020stv}, showing clear advantages of muon colliders.  Here we mainly compare our results  with $VV\rightarrow hh$ at muon colliders.  In ref.~\cite{Costantini:2020stv}, the projective limits on $c_6$ and $c_{\Phi_1}$ at muon colliders with $\sqrt{s}=14$ TeV and the integrated luminosity of $20 \ \text{ab}^{-1}$ are obtained by combining $VV\rightarrow hh$ and $VV\rightarrow hhh$.  The results are
\begin{equation}
c_6\in [-0.66, 0.23],\qquad  c_{\Phi_1}\in [-0.17, 0.30]
\end{equation}
at $95\%$ for $\Lambda = 1 $ TeV. This is  close to our similar results for both $VV\rightarrow W^+_LW^-_Lh$ and $VV\rightarrow hhh$, as can be seen in~\autoref{tab:test1} and~\autoref{tab:test2}, indicating that  constraints on Higgs self-couplings from $2\rightarrow 3$ VBS might be comparable to $VV\rightarrow hh$ at muon colliders.  Of course, more careful and in-depth studies are obviously needed.

\section{Conclusions}
\label{sec:conclusion}

Measuring Higgs self-couplings is a crucial task for the future collider experiments. It will uncover the nature of the discovered SM-like Higgs bosn, the origin of EWSB, and the shape of the Higgs potential, etc.
In this work,
we studied the $2\rightarrow 3$ VBS processes under the framework of SMEFT, using $W^{\pm}_LW^{\pm}_L\rightarrow W^{\pm}_LW^{\pm}_Lh$ and $W^+_LW^-_L\rightarrow hhh$ as examples.

First, the behavior of the amplitudes for those processes at high energy is analysed using GET. We found that compared with SM contributions, the BSM contribution will be enhanced at high energy as $\frac{\mathcal{A}_{\rm BSM}}{\mathcal{A}_{\rm SM}}\sim\frac{E^2}{\Lambda^2}$. Physically, this behavior comes from the combination of the following two factors: SM amplitudes are suppressed by the energy in the propagators; BSM amplitudes can stay constant due to contact vertex from $\mathcal{O}_6$ or increase with energy due to the momentum dependence in $\mathcal{O}_{\Phi_1}$. By numerically calculating the cross section for these $2\to3$ processes, we showed the sensitivities of these processes to $c_6$ and $c_{\Phi_1}$ in~\autoref{fig:C6-1} to~\autoref{fig:cs_ratio_cphi}.

Second, the processes $pp\rightarrow jj W^{\pm}_LW^{\pm}_Lh$ and $pp\rightarrow jjhhh$ at pp colliders and $\mu^+ \mu^-\rightarrow\nu_{\mu}\overline{\nu}_{\mu} W^+_LW^-_Lh$ and $\mu^+ \mu^-\rightarrow\nu_{\mu}\overline{\nu}_{\mu}hhh$ at muon colliders are simulated with various energy benchmark points.
However, certain $p_T$ cuts in the phase space of final state particles are needed for $pp\rightarrow jj W^{\pm}_LW^{\pm}_Lh$ and $\mu^+ \mu^-\rightarrow\nu_{\mu}\overline{\nu}_{\mu} W^+_LW^-_Lh$ processes, in order to reduce the SM cross sections enhanced by Sudakov logarithms from collinear divergences. We study in details on how cross sections change with energies and with the Wilson coefficients $c_6$ and $c_{\Phi_1}$ in~\autoref{fig:cs_wwh} to~\autoref{fig:ratio_hhh}.

Based on the simulation, we estimated the allowed regions of $c_6$ and $c_{\Phi_1}$ at different c.m. energies and types of colliders assuming that signal events are already extracted from relevant SM backgrounds for processes in~\autoref{proc:ee} and~\autoref{proc:pp}.
Cross sections of these processes are generally tiny at HL-LHC, HE-LHC and CLIC, which makes them very challenging to  explore.
On the other hand, we find at the future 100 TeV pp colliders or high energy muon colliders, these VBS processes are good processes  to measure the Higgs self-couplings, parameterized by $c_6$ and $c_{\Phi_1}$ in SMEFT. The allowed 1-$\sigma$ and 2-$\sigma$ regions for $c_6$ and $c_{\Phi_1}$ are obtained based on simple event counting procedure and are summarized in~\autoref{fig:allowed_region_c6_cphi}.
 Optimistically, we expect  $-0.23<c_6<0.18, -0.07<c_{\Phi_1} < 0.04$ at $1\sigma$ can be reached in the future and $c_{\Phi_1}$ is more restrictive than $c_6$.

 We find  $WWh$ process is as important as the more widely studied triple Higgs production ($hhh$)~\cite{Chen:2015gva,Fuks:2015hna,Dicus:2016rpf,Fuks:2017zkg,Agrawal:2017cbs,Belyaev:2018fky,Maltoni:2018ttu,Kilian:2018bhs,Agrawal:2019bpm,Papaefstathiou:2019ofh,deFlorian:2019app,Chiesa:2020awd,Gonzalez-Lopez:2020lpd} in the measurement of  Higgs self-couplings. Our analysis is only preliminary, as we aim to give an overall picture and qualitative conclusions. A more careful analysis that takes into account of decay products, relevant SM background  and detector effects is obviously needed. Moreover, we only studied a partial list of all $2\rightarrow 3$ VBS processes, with $W^{\pm}W^{\pm}h/hhh$ ($W^{+}W^{-}h/hhh$) as final states at hadron (lepton) colliders. We will devote a more complete survey for all $2\rightarrow 3$ VBS processes in measuring Higgs self-couplings in future work.

Furthermore, the enhancement of the amplitude at high energy in the presence of BSM physics due to contact scalar vertices is not limited to the case of $2\rightarrow 3$ VBS processes. For example, the analysis in this paper can also be applied to $2\rightarrow 4$ VBS processes which have the same energy increase behavior. Hence, they can also be used to measure Higgs self-coupling at future high energy colliders.
It would be interesting to explore further on this direction.

\begin{acknowledgments}
We would like to thank helpful discussions with Yang Ma, Tong Li. CTL is supported by KIAS Individual Grant, No.PG075301 at Korea Institute for Advanced Study. JMC is supported by Fundamental Research Funds for the Central Universities of China NO.11620330.
\end{acknowledgments}


\appendix

\section{Full Cross Sections Summing Over Polarization Vector Bosons }\label{Sec:Appendix-2}

In this appendix, we sum over polarizations for final state $W^{\pm}$s of $\mu^+ \mu^-\rightarrow \nu_{\mu}\bar{\nu}_\mu W^+ W^- h$ and $p p \rightarrow j j W^+W^- h$ processes in~\autoref{tab:eevvw+w-h} and~\autoref{tab:ppjjw+w-h}, respectively. Compared with~\autoref{tab:eevvw+0w-0h} and~\autoref{tab:ppjjw+0w-0h}, the relevant cross sections can be larger by one order, but the relative deviations from SM ones are not obvious, especially for the $c_6$ contributions.
Therefore, a precise way to distinguish $W^{\pm}_L$ from $W^{\pm}_T$ is important to extract effects from the ${\mathcal O}_6$ operator. In this work, we assume $W^{\pm}_L$ can already be successfully separated from $W^{\pm}_T$ experimentally. More details about the method to distinguish $W^{\pm}_L$ from $W^{\pm}_T$ can be found in~\cite{Ballestrero:2017bxn,Ballestrero:2020qgv,De:2020iwq,Grossi:2020orx,Kim:2021gtv}.

\begin{table}[!tb]
\centering
  \begin{tabular}{ |c|c|c|c|c|c|}
 \hline
 \multicolumn{6}{|c|}{Cross sections (pb) for $\mu^+ \mu^-\rightarrow \nu_\mu\bar{\nu}_\mu W^+ W^- h $ with $c_{\Phi_1}=0$  }\\
 \hline
$c_6$  &  $-2$  &  $-1$  &  $0$  &  $1$  &  $2$  \\
 \hline
1 TeV  & $2.87\times 10^{-8}$ & $2.51\times 10^{-8}$ & $2.37\times 10^{-8}$ & $2.44\times 10^{-8}$ & $2.73\times 10^{-8}$ \\
 \hline
3 TeV  & $2.04\times 10^{-5}$ & $1.90\times 10^{-5}$ &   $1.85\times 10^{-5}$ & $1.88\times 10^{-5}$ & $2.01\times 10^{-5}$ \\
 \hline
5 TeV  & $8.18\times 10^{-5}$ & $7.74\times 10^{-5}$ &  $7.60\times 10^{-5}$ & $7.72\times 10^{-5}$ & $8.07\times 10^{-5}$ \\
 \hline
10 TeV & $3.16\times 10^{-4}$ & $3.02\times 10^{-4}$ &  $3.00\times 10^{-4}$ & $3.02\times 10^{-4}$ & $3.13\times 10^{-4}$ \\
 \hline
14 TeV & $5.29\times 10^{-4}$ & $5.12\times 10^{-4}$ &  $5.03\times 10^{-4}$ & $5.06\times 10^{-4}$ & $5.29\times 10^{-4}$ \\
 \hline
30 TeV & $1.38\times 10^{-3}$ & $1.31\times 10^{-3}$ &  $1.31\times 10^{-3}$ & $1.33\times 10^{-3}$ & $1.36\times 10^{-3}$  \\
 \hline
\end{tabular}

 \

 \

  \begin{tabular}{ |c|c|c|c|c|c|}
 \hline
 \multicolumn{6}{|c|}{Cross sections (pb) for $\mu^+ \mu^-\rightarrow \nu_\mu\bar{\nu}_\mu W^+ W^- h $ with $c_6=0$ }\\
 \hline
$c_{\Phi_1}$  &  $-2$  &  $-1$  &  $0$  &  $1$  &  $2$  \\
 \hline
1 TeV  & $5.99\times 10^{-6}$ & $6.85\times 10^{-6}$ & $7.97\times 10^{-6}$ & $9.40\times 10^{-6}$ & $1.11\times 10^{-5}$ \\
 \hline
3 TeV  & $2.39\times 10^{-4}$ & $2.51\times 10^{-4}$ &  $2.81\times 10^{-4}$ & $3.29\times 10^{-4}$ & $3.90\times 10^{-4}$ \\
 \hline
5 TeV  & $7.15\times 10^{-4}$ & $7.00\times 10^{-4}$ &  $7.58\times 10^{-4}$ & $8.96\times 10^{-4}$ & $1.11\times 10^{-3}$ \\
 \hline
10 TeV & $2.66\times 10^{-3}$ & $2.12\times 10^{-3}$ &  $2.12\times 10^{-3}$ & $2.66\times 10^{-3}$ & $3.69\times 10^{-3}$ \\
 \hline
14 TeV & $5.07\times 10^{-3}$ & $3.48\times 10^{-3}$ &  $3.15\times 10^{-3}$ & $4.22\times 10^{-3}$ & $6.50\times 10^{-3}$ \\
 \hline
30 TeV & $2.49\times 10^{-2}$ & $1.10\times 10^{-2}$ &  $6.81\times 10^{-3}$ & $1.25\times 10^{-2}$ & $2.76\times 10^{-2}$ \\
 \hline
\end{tabular}
 \caption{ The $\mu^+ \mu^-\rightarrow \nu_{\mu}\bar{\nu}_\mu W^+ W^- h$ process with summing over polarizations for final $W^{\pm}$, with different c.m. energies, varies with $c_6$, $c_{\Phi_1}=0$ (upper table) and varies with $c_{\Phi_1}$, $c_6=0$ (lower table). Same cuts listed in~\autoref{tab:cuts} are applied.
 }
 \label{tab:eevvw+w-h}
\end{table}

\begin{table}[!tb]
  \centering
 \begin{tabular}{ |c|c|c|c|c|c|}
 \hline
 \multicolumn{6}{|c|}{Cross sections (pb) for $pp\rightarrow jj W^{\pm}W^{\pm}h $ with $c_{\Phi_1}=0$ }\\
 \hline
$c_6$  &  $-2$  &  $-1$  &  $0$  &  $1$  &  $2$  \\
 \hline
14 TeV  & $1.39\times 10^{-5}$ & $1.37\times 10^{-5}$ &  $1.36\times 10^{-5}$ & $1.37\times 10^{-5}$ & $1.42\times 10^{-5}$ \\
 \hline
27 TeV  & $9.11\times 10^{-5}$ & $8.93\times 10^{-5}$ &  $8.91\times 10^{-5}$ & $8.98\times 10^{-5}$ & $9.15\times 10^{-5}$ \\
 \hline
100 TeV & $1.13\times 10^{-3}$ & $1.11\times 10^{-3}$ &  $1.11\times 10^{-3}$ & $1.12\times 10^{-3}$ & $1.13\times 10^{-3}$ \\
 \hline
\end{tabular}

 \

 \

 \begin{tabular}{ |c|c|c|c|c|c|}
 \hline
 \multicolumn{6}{|c|}{Cross sections (pb) for $pp\rightarrow jj W^{\pm}W^{\pm}h $ with $c_6=0$}\\
 \hline
$c_{\Phi_1}$  &  $-2$  &  $-1$  &  $0$  &  $1$  &  $2$  \\
 \hline
14 TeV  & $2.26\times 10^{-4}$ & $2.51\times 10^{-4}$ &  $2.83\times 10^{-4}$ & $3.21\times 10^{-4}$ & $3.63\times 10^{-4}$ \\
 \hline
27 TeV  & $1.04\times 10^{-3}$ & $1.14\times 10^{-3}$ &  $1.27\times 10^{-3}$ & $1.46\times 10^{-3}$ & $1.67\times 10^{-3}$ \\
 \hline
100 TeV & $9.15\times 10^{-3}$ & $9.48\times 10^{-3}$ &  $1.04\times 10^{-2}$ & $1.20\times 10^{-2}$ & $1.42\times 10^{-2}$ \\
 \hline
\end{tabular}
 \caption{ The same as~\autoref{tab:eevvw+w-h} for the $p p \rightarrow j j W^+W^- h$ process. Same cuts listed in~\autoref{tab:cuts} are applied. }
 \label{tab:ppjjw+w-h}
\end{table}

 \null\newpage


\bibliographystyle{JHEP}
\bibliography{references}

\end{document}